%% file: cowan_sussp_preprint.tex
\documentclass[graybox]{svmult}

\usepackage{mathptmx}       
\usepackage{helvet}         
\usepackage{courier}        
\usepackage{type1cm}        
\usepackage{makeidx}         
\usepackage{graphicx}        
\usepackage{multicol}        
\usepackage[bottom]{footmisc}

\makeindex             

\def\bvec#1{\ifmmode
\mathchoice{\mbox{\boldmath$\displaystyle\bf#1$}}
{\mbox{\boldmath$\textstyle\bf#1$}}
{\mbox{\boldmath$\scriptstyle\bf#1$}}
{\mbox{\boldmath$\scriptscriptstyle\bf#1$}}\else
{\mbox{\boldmath$\bf#1$}}\fi}

\begin{document}

\title*{Statistics for Searches at the LHC}
\author{Glen Cowan} 
%
%
\maketitle

\vspace{-4 cm}
\noindent Physics Department, Royal Holloway, University of London, \\
\noindent Egham, Surrey, TW20 0EX, UK
\vspace{3.5 cm}

\abstract*{These lectures\footnote{Lectures presented at the 69th
    Scottish Universities Summer School in Physics, St.~Andrews,
    August-September 2012.} describe several topics in statistical
data analysis as used in High Energy Physics.  They focus on areas
most relevant to analyses at the LHC that search for new physical
phenomena, including statistical tests for discovery and exclusion
limits.  Particular attention is payed to the treatment of systematic
uncertainties through nuisance parameters.}

\abstract{These lectures\footnote{Lectures presented at the 69th
    Scottish Universities Summer School in Physics, St.~Andrews,
    August-September 2012.} describe several topics in statistical
data analysis as used in High Energy Physics.  They focus on areas
most relevant to analyses at the LHC that search for new physical
phenomena, including statistical tests for discovery and exclusion
limits.  Particular attention is payed to the treatment of systematic
uncertainties through nuisance parameters.}

\section{Introduction}
\label{sec:intro}

The primary goal of data analysis in High Energy Physics (HEP) is to
test our understanding of particle interactions and in doing so to
search for phenomena that go beyond the existing framework of the
Standard Model.  These lectures describe some of the basic statistical
tools needed to do this.

Despite efforts to make the lectures self contained, some familiarity
with basic ideas of statistical data analysis is assumed.
Introductions to the subject can be found, for example, in the reviews
of the Particle Data Group~\cite{bib:PDG} or in the
texts~\cite{bib:Cowan98,bib:Lyons,bib:Barlow,bib:James,bib:Brandt}.

Brief reviews are given of probability in Sec.~\ref{sec:prob} and
frequentist hypothesis tests in Secs.~\ref{sec:hyptest} and
\ref{sec:critreg}.  These are applied to establishing discovery and
setting limits (Sec.~\ref{sec:disclim}) and are extended using the
profile likelihood ratio (Sec.~\ref{sec:pl}), from which one can
construct unified intervals (Sec.~\ref{sec:fc}).  Bayesian limits are
discussed in Sec.~\ref{sec:bayeslim} and all of the methods for limits
are illustrated using the example of a Poisson counting experiment in
Sec.~\ref{sec:poislim}.  Application of the standard tools for
discovery and limits leads to a number of difficulties, such as
exclusion of models to which one has no sensitivity
(Sec.~\ref{sec:spurious}) and the look-elsewhere effect
(Sec.~\ref{sec:lee}).  Section~\ref{sec:higgs} illustrates how the
methods have been applied in the search for the Higgs boson.  In
Sec.~\ref{sec:why5} we examine why one traditionally requires
five-sigma significance to claim a discovery and finally some
conclusions are drawn in Sec.~\ref{sec:conclusions}.  The lectures as
presented at SUSSP also included material on unfolding or
deconvolution of measured distributions which is not included here but
can be found in Ref.~\cite{bib:gdcunfold} and Chapter 11 of
Ref.~\cite{bib:Cowan98}.

\section{Review of probability}
\label{sec:prob}

When analyzing data in particle physics one invariably encounters
uncertainty, at the very least coming from the intrinsically random
nature of quantum mechanics.  These uncertainties can be quantified
using probability, which was defined by Kolmogorov
\cite{bib:Kolmogorov33} using the language of set theory.  Suppose a
set $S$ contains elements that can form subsets $A$, $B$, $\ldots$.
As an example, the elements may represent possible outcomes of a
measurement but here we are being abstract and we do not need to
insist at this stage on a particular meaning.  The three axioms of
Kolmogorov can be stated as

\begin{enumerate}

\item For all $A \subset S$, there is a real-valued function $P$, the
  probability, with $P(A) \ge 0$;

\item $P(S) = 1$;

\item If $A \cap B = \emptyset$, then $P(A \cup B) = P(A) + P(B)$.

\end{enumerate}

In addition we define the conditional probability of $A$ given $B$
(for $P(B) \ne 0$) as

\begin{equation}
\label{eq:condprob}
P(A|B) = \frac{P(A \cap B)}{P(B)} \;.
\end{equation}

\noindent From these statements we can derive the familiar properties
of probability.  They do not, however, provide a complete recipe for
assigning numerical values to probabilities nor do tell us what these
values mean.

Of the possible ways to interpret a probability, the one most commonly
found in the physical sciences is as a limiting frequency.  That is,
we interpret the elements of the sample space as possible outcomes of
a measurement, and we take $P(A)$ to mean the fraction of times that
the outcome is in the subset $A$ in the limit where we repeat the
measurement an infinite number of times under ``identical''
conditions:

\begin{equation}
\label{eq:freqprob}
P(A) = \lim_{n  \rightarrow \infty} \frac{\mbox{times outcome is in $A$}}{n} \;.
\end{equation}

Use of probability in this way leads to what is called the {\it
  frequentist} approach to statistics.  Probabilities are only
associated with outcomes of repeatable observations, not to
hypothetical statements such as ``supersymmetry is true''.  Such a
statement is either true or false, and this will not change upon
repetition of any experiment.

Whether SUSY is true or false is nevertheless uncertain and we can
quantify this using probability as well.  To define what is called
{\it subjective probability} one interprets the elements of the set
$S$ as {\it hypotheses}, i.e., statements that are either true or
false, and one defines

\begin{equation}
P(A) = \mbox{degree of belief that } A \mbox{ is true.}
\end{equation}

\noindent Use of subjective probability leads to what is called {\it
  Bayesian statistics}, owing to its important use of Bayes' theorem
described below.

Regardless of its interpretation, any quantity that satisfies the
axioms of probability must obey Bayes' theorem, which states

\begin{equation}
\label{eq:bayesthm}
P(A|B) = \frac{P(B|A) P(A)}{P(B)} \;.
\end{equation}

\noindent This can be derived from the definition of conditional
probability (\ref{eq:condprob}), which we can write as $P(A \cap B) =
P(B) P(A |B)$, or equivalently by changing labels as $P(B \cap A) =
P(A) P(A |B)$.  These two probabilities are equal, however, because $A
\cap B$ and $B \cap A$ refer to the same subset.  Equating them leads
directly to Eq.~(\ref{eq:bayesthm}).

In Bayes' theorem (\ref{eq:bayesthm}) the condition $B$ represents a
restriction imposed on the sample space $S$ such that anything outside
of $B$ is not considered.  If the sample space $S$ can be expressed as
the union of some disjoint subsets $A_i$, $i = 1, 2, \ldots$, then the
factor $P(B)$ appearing in the denominator can be written $P(B) =
\sum_i P(B|A_i) P(A_i)$ so that Bayes' theorem takes on the form

\begin{equation}
\label{eq:bayesthm2}
P(A|B) = \frac{P(B|A) P(A)}{\sum_i P(B|A_i) P(A_i)} \;.
\end{equation}

In {\it Bayesian} (as opposed to frequentist) statistics, one uses
subjective probability to describe one's degree of belief in a given
theory or hypothesis.  The denominator in Eq.~(\ref{eq:bayesthm2}) can
be regarded as a constant of proportionality and therefore Bayes'
theorem can be written as

\begin{equation}
\label{eq:bayesthm3}
P(\hbox{theory}|\hbox{data}) \propto 
P(\hbox{data}|\hbox{theory}) P(\hbox{theory}) \;,
\end{equation}

\noindent where ``theory'' represents some hypothesis and ``data'' is
the outcome of the experiment.  Here $P(\hbox{theory})$ is the {\it
  prior} probability for the theory, which reflects the experimenter's
degree of belief before carrying out the measurement, and
$P(\hbox{data}|\hbox{theory})$ is the probability to have gotten the
data actually obtained, given the theory, which is also called the
{\it likelihood}.

Bayesian statistics provides no fundamental rule for obtaining the
prior probability; in general this is subjective and may depend on
previous measurements, theoretical prejudices, etc. Once this has been
specified, however, Eq.~(\ref{eq:bayesthm3}) tells how the probability
for the theory must be modified in the light of the new data to give
the {\it posterior} probability, $P(\hbox{theory}|\hbox{data})$.  As
Eq.~(\ref{eq:bayesthm3}) is stated as a proportionality, the
probability must be normalized by summing (or integrating) over all
possible hypotheses.

\section{Hypothesis tests}
\label{sec:hyptest}

One of the fundamental tasks in a statistical analysis is to test
whether the predictions of a given model are in agreement with the
observed data.  Here we will use $\bvec{x}$ to denote the outcome of a
measurement; it could represent a single quantity or a collection of
values.  A hypothesis $H$ means a statement for the probability to find
the data $\bvec{x}$ (or if $\bvec{x}$ includes continuous variables,
$H$ specifies a probability density function or pdf).  We will write
$P(\bvec{x} | H)$ for the probability to find data $\bvec{x}$ under
assumption of the hypothesis $H$.

Consider a hypothesis $H_0$ that we want to test (we will often call
this the ``null'' hypothesis) and an alternative hypothesis $H_1$.  In
frequentist statistics one defines a {\it test} of $H_0$ by specifying
a subset of the data space called the {\it critical region}, $w$, such
that the probability to observe the data there satisfies

\begin{equation}
\label{eq:critreg}
  P(\bvec{x} \in w | H_0) \le \alpha \;.
\end{equation}

\noindent Here $\alpha$ is a constant specified before carrying out
the test, usually set by convention to a small value such as 5\%.  For
continuous data, one takes the relation above as an equality.  If the
data are discrete, such as a number of events, then there may not
exist any subset of the data values whose summed probability is
exactly equal to $\alpha$, so one takes the critical region to have a
probability up to $\alpha$.  The critical region $w$ defines the test.
If the data are observed in $w$ then the hypothesis $H_0$ is rejected.

Up to this point the sole defining property of the test is
Eq.~(\ref{eq:critreg}), which states that the probability to find the
data in the critical region is not more than $\alpha$.  But there are
in general many if not an infinite number of possible subsets of the
data space that satisfy this criterion, and it is not clear which
should be taken as the critical region.  This is where the alternative
hypothesis $H_1$ comes into play.  One would like the critical region
to be chosen such that there is as high a probability as possible to
find the data there if the alternative is true, while having only the
fixed probability $\alpha$ assuming $H_0$, as illustrated
schematically in Fig.~\ref{fig:criticalRegion}.

\setlength{\unitlength}{1.0 cm}
\renewcommand{\baselinestretch}{0.8}
\begin{figure}[htbp]
\begin{picture}(10.0,5.5)
\put(0,-0.2){\includegraphics[width=0.55\textwidth]{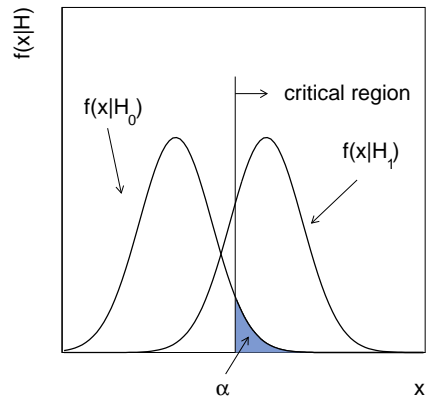}}
\put(6.7,0.2){\makebox(5,4)[b]{\begin{minipage}[b]{5cm}
\protect\caption{{\small Illustration of the critical region
of a statistical test (see text).}
\protect\label{fig:criticalRegion}}
\end{minipage}}}
\end{picture}
\end{figure}
\renewcommand{\baselinestretch}{1}
\small\normalsize

Rejecting the hypothesis $H_0$ if it is in fact true is called a type
I error.  By construction the probability for this to occur is the
size of the test, $\alpha$.  If on the other hand we do not reject
$H_0$ but we should have, because the alternative $H_1$ was true, then
this is called a type II error.  The probability to reject the null
hypothesis if the alternative $H_1$ is true is called the {\it power}
of the test with respect to $H_1$, which is one minus the probability
of a type II error.

%

A {\it significance test} of a hypothesis $H$ is closely related to
the tests described above.  Suppose a measurement results in data
$\vec{x}$ (a single number or a collection of many values) for which
the hypothesis $H$ predicts the probability $P(\bvec{x}|H)$.  We
observe a single instance of $\bvec{x}$, say, $\bvec{x}_{\rm obs}$,
and we want to quantify the level of agreement between this outcome
and the predictions of $H$.

To do this the analyst must specify what possible data values would
constitute a level of incompatibility with $H$ that is equal to are
greater than that between $H$ and the observed data $\bvec{x}_{\rm
  obs}$.  Once this is given, then one computes the $p$-value of $H$
as the probability, under assumption of $H$, to find data in this
region of equal or greater incompatibility.

When computing the $p$-value there is clearly some ambiguity as to
what data values constitute greater incompatibility with $H$ than
others.  When we say that a given $\bvec{x}$ has less compatibility
with $H$, we imply that it has more compatibility with some
alternative hypothesis.  This is analogous to the ambiguity we
encountered in determining the critical region of a test.

We can see this connection more directly by using the $p$-value to
specify the critical region for a test of $H_0$ of size $\alpha$ as
the set of data values that would have a $p$-value less than or equal
to $\alpha$.  The resulting test will have a certain power with
respect to any given alternative $H_1$, although these may not have
been used explicitly when constructing the $p$-value.

In a frequentist test, we {\it reject} $H_0$ if the data are found in
the critical region, or equivalently, if the $p$-value of $H_0$ is
found less or equal to $\alpha$.  Despite this language, it is not
necessarily true that we would would then believe $H_0$ to be false.
To make this assertion we should quantify our degree of belief about
$H_0$ using subjective probability as described above, and it must be
computed using Bayes' theorem:

\begin{equation}
\label{eq:bayesH}
P(H_0 | \bvec{x} ) = \frac{ P(\bvec{x} | H_0) \pi (H_0) }
{\sum_i P(\bvec{x} | H_i) \pi(H_i) } \;.
\end{equation}

\noindent As always, the posterior $P(H_0 | \bvec{x} )$ is
proportional to the prior $\pi(H_0)$, and this would need to be
specified if we want to express our degree of belief that the
hypothesis is true.

It is also important to note that the $p$-value of a hypothesis $H_0$
is not the same as the probability (\ref{eq:bayesH}) that it is true,
but rather the probability, under assumption of $H_0$, to find data
with at least as much incompatibility with $H_0$ as the data actually
found.  The $p$-value thus does not depend on prior probabilities.


For most of these lectures we will stay within the frequentist
framework.  The result of our analysis will be a $p$-value for the
different models considered.  If this is less than some specified
value $\alpha$, we reject the model.


Often the $p$-value is translated into an equivalent quantity called
the {\it significance}, $Z$, defined by

\begin{equation}
\label{eq:significance}
Z = \Phi^{-1}(1 -p) \;.
\end{equation}

\noindent Here $\Phi$ is the cumulative standard Gaussian distribution
(zero mean, unit variance) and $\Phi^{-1}$ is its inverse function,
also called the {\it quantile} of the standard Gaussian.  The
definition of significance is illustrated in
Fig.~\ref{fig:significance}(a) and the significance versus $p$-value
is shown in Fig.~\ref{fig:significance}(b).  Often a significance of
$Z = 5$ is used as the threshold for claiming discovery of a new
signal process.  This corresponds to a very low $p$-value of $2.9
\times 10^{-7}$ for the no-signal hypothesis.  The rationale for using
such a low threshold is discussed further in Sec.~\ref{sec:why5}.

Although we can simply take Eq.~(\ref{eq:significance}) as our
defining relation for $Z$, it is useful to compare to the case of
measuring a quantity $x$ that follows a Gaussian distribution with
unknown mean $\mu$.  Suppose we want to test the hypothesis $\mu=0$
against the alternative $\mu > 0$.  In this case we would take the
critical region of the test to contain values of $x$ greater than a
certain threshold, or equivalently, we would define the $p$-value to
be the probability to find $x$ as large as we found or larger.  In
this case the significance $Z$ is simply the value of $x$ observed,
measured in units of its standard deviation $\sigma$.  For this reason
one often refers to finding a significance $Z$ of, say, 2.0 as a
$two$-sigma effect.

\setlength{\unitlength}{1.0 cm}
\renewcommand{\baselinestretch}{0.9}
\begin{figure}[htbp]
\begin{picture}(10.0,5.)
\put(-0.5,0){\includegraphics[width=0.45\textwidth]{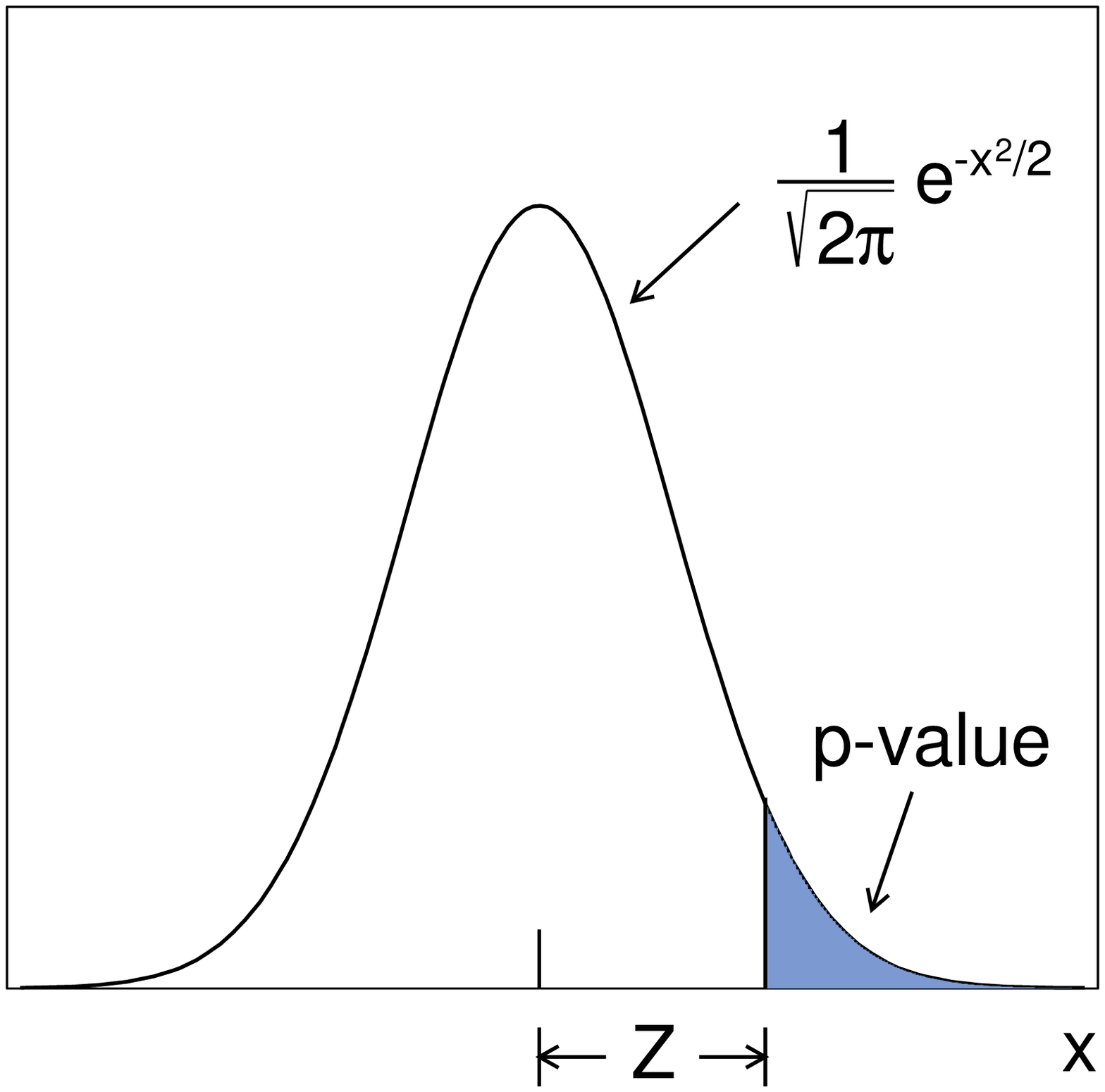}}
\put(6.,0){\includegraphics[width=0.45\textwidth]{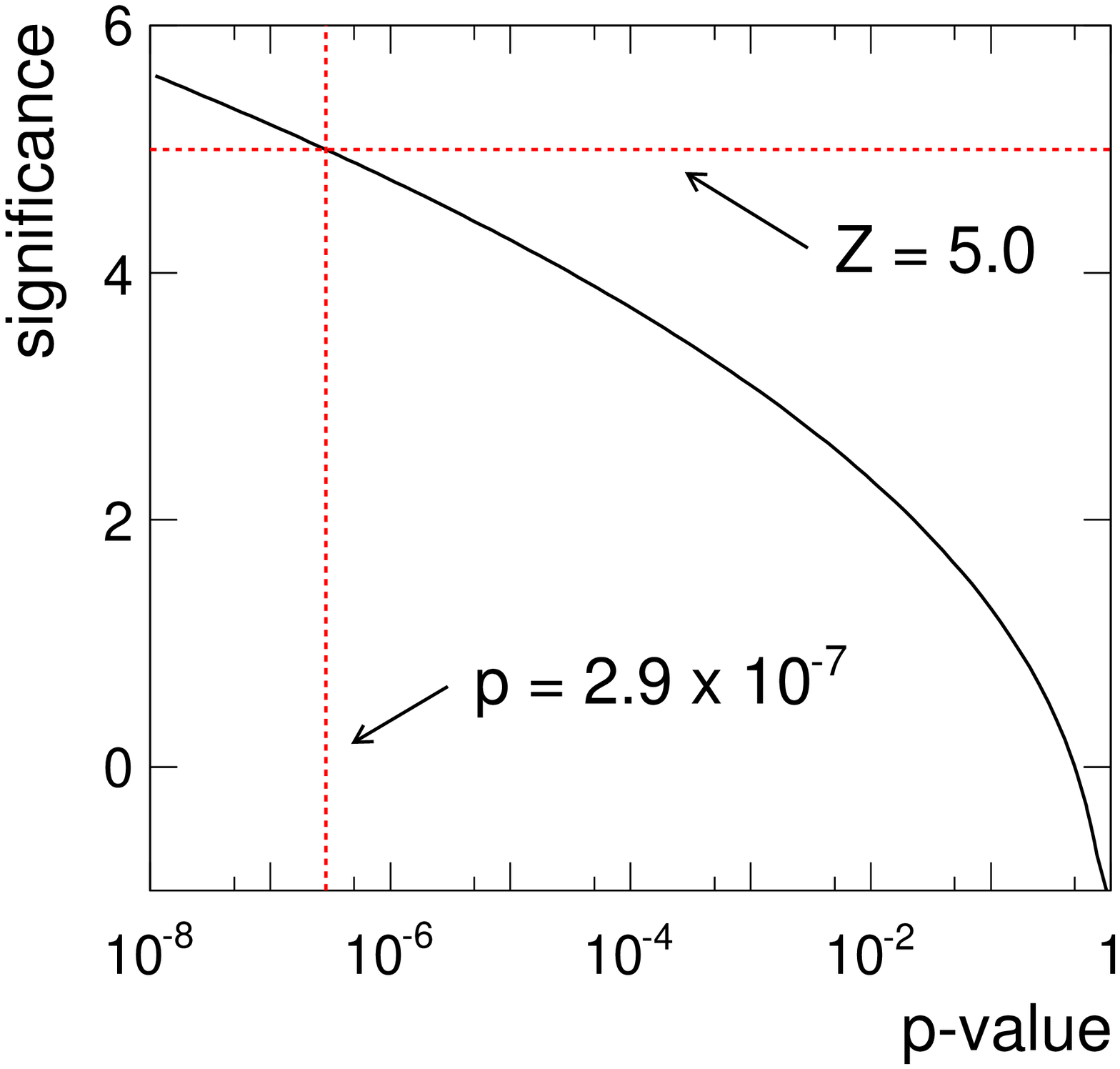}}
\put(4.7,4.5){(a)}
\put(11.3,4.5){(b)}
\end{picture}
\caption{\small (a) Illustration of the definition of significance
$Z$ and (b) the significance as function of the $p$-value.}
\label{fig:significance}
\end{figure}
\renewcommand{\baselinestretch}{1}
\small\normalsize

\section{Choice of critical region, test statistics}
\label{sec:critreg}

We now examine more closely the question of how best to define the
critical region of a test and for this suppose we want to select a
sample of events of a desired type (signal, denoted $s$) and reject
others that we do not want (background, $b$).  That is, for each event
we will measure some set of quantities $\bvec{x}$, which could
represent different kinematic variables such as the missing energy,
number of jets, number of muons, and so forth.  Then for each event
carry out a test of the background hypothesis, and if this is rejected
it means we select the event as a candidate signal event.

Suppose that the $s$ and $b$ hypotheses imply probabilities for the
data of $P(\bvec{x}|s)$ and $P(\bvec{x}|b)$, respectively.
Figures~\ref{fig:scatter}(a -- c) show these densities for two
components of the data space along with possible boundaries for the
critical region.

\setlength{\unitlength}{1.0 cm}
\renewcommand{\baselinestretch}{0.9}
\begin{figure}[htbp]
\begin{picture}(10.0,3.5)
\put(0.,-0.2){\includegraphics[width=0.333\textwidth]{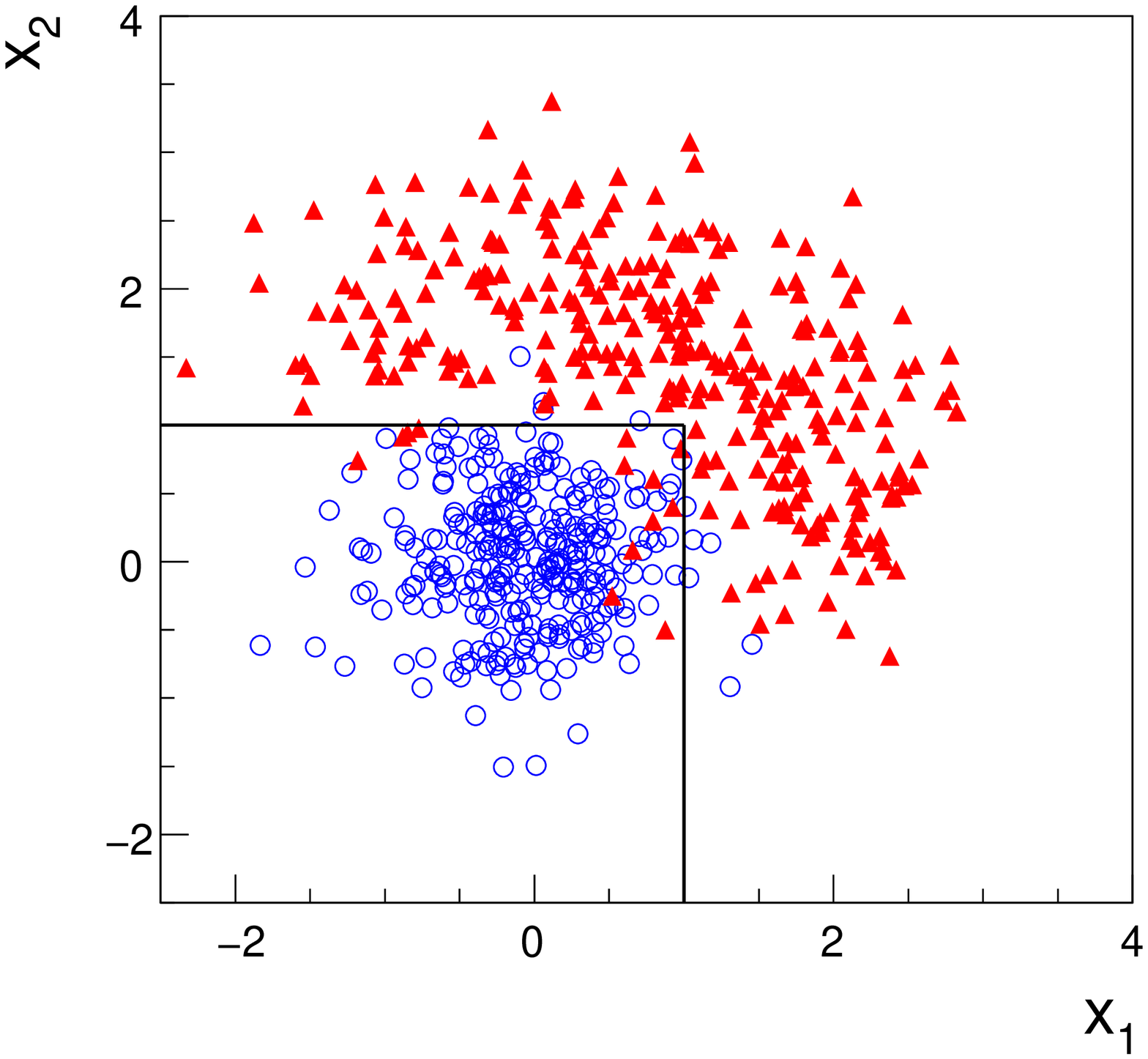}}
\put(4,-0.2){\includegraphics[width=0.333\textwidth]{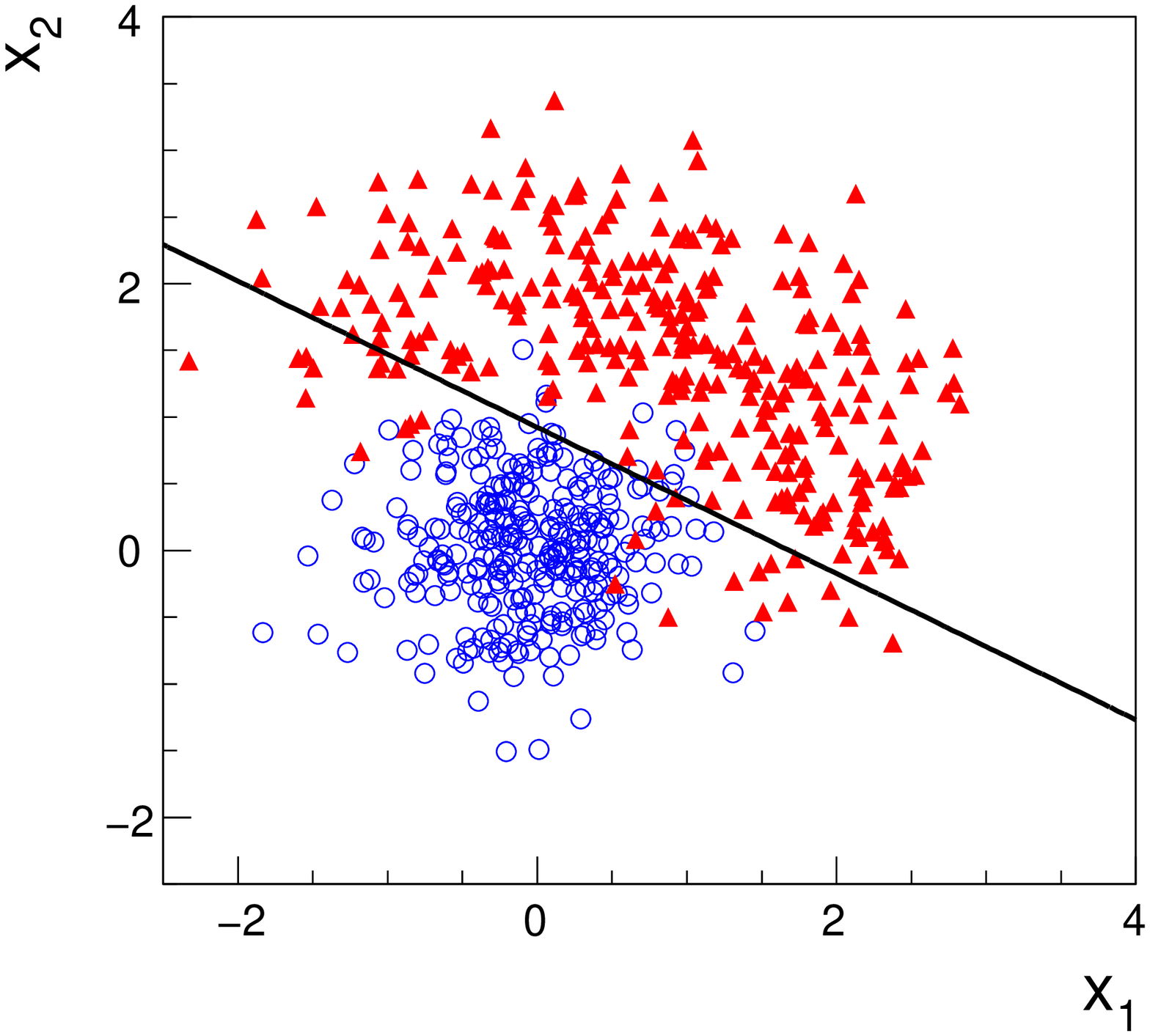}}
\put(8,-0.2){\includegraphics[width=0.333\textwidth]{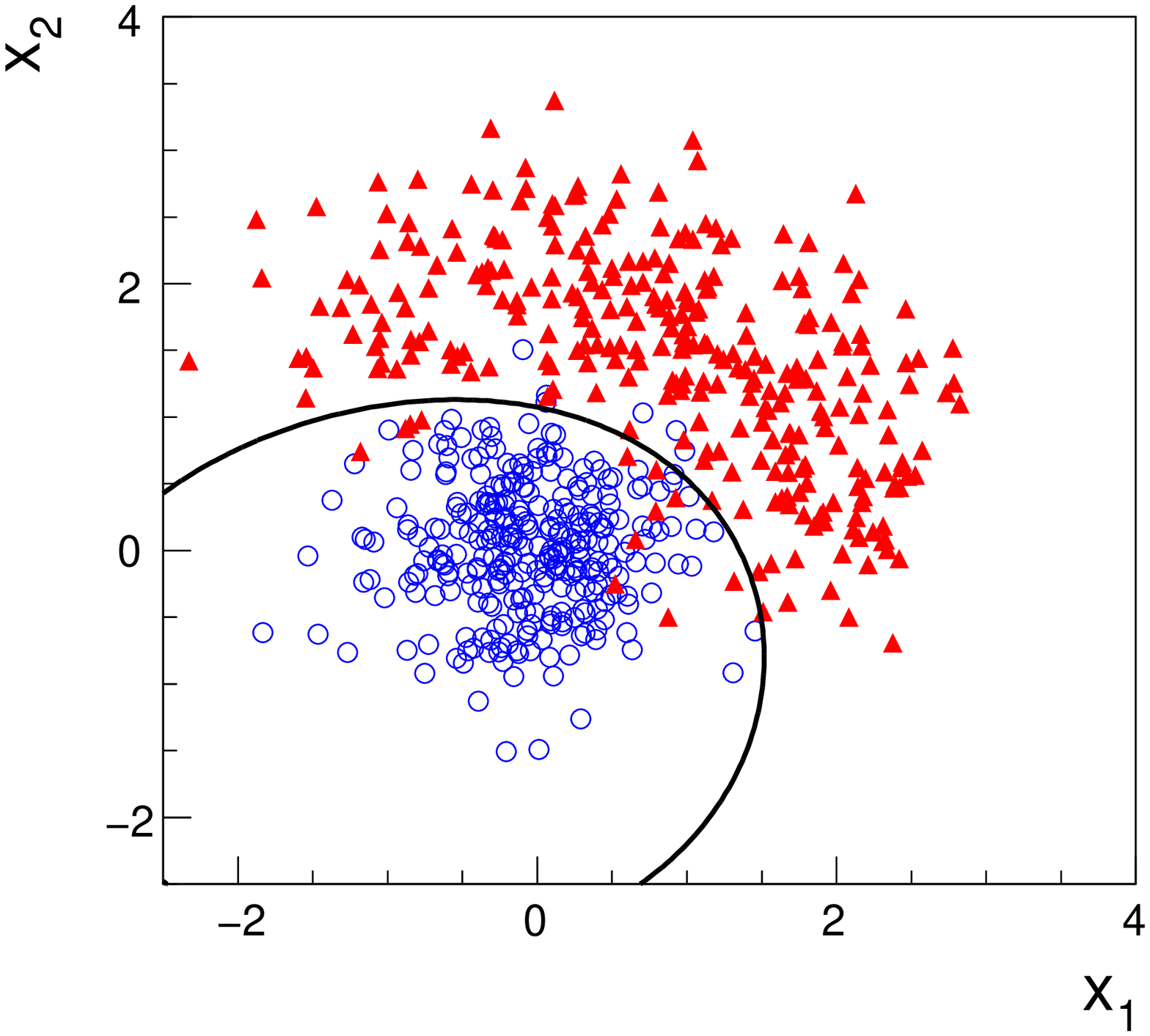}}
\put(3.2,2.9){(a)}
\put(7.2,2.9){(b)}
\put(11.2,2.9){(c)}
\end{picture}
\caption{\small Scatter plots of two variables corresponding to two
  hypotheses: background ($H_0$) and signal ($H_1$).  The critical
  region for a test of $H_0$ could be based, e.g., on (a) cuts, (b) a
  linear boundary, (c) a nonlinear boundary.}
\label{fig:scatter}
\end{figure}
\renewcommand{\baselinestretch}{1}
\small\normalsize

Figure~\ref{fig:scatter}(a) represents what is commonly called the
`cut-based' approach.  One selects signal events by requiring $x_1 <
c_1$ and $x_2 < c_2$ for some suitably chosen cut values $c_1$ and
$c_2$.  If $x_1$ and $x_2$ represent quantities for which one has some
intuitive understanding, then this can help guide one's choice of the
cut values.

Another possible decision boundary is made with a diagonal cut as
shown in Fig.~\ref{fig:scatter}(b).  One can show that for certain
problems a linear boundary has optimal properties, but in
the example here, because of the curved nature of the distributions,
neither the cut-based nor the linear solution is as good as the
nonlinear boundary shown in Fig.~\ref{fig:scatter}(c).

The decision boundary is a surface in the $n$-dimensional space of
input variables, which can be represented by an equation of the form
$y(\vec{x}) = y_{\rm cut}$, where $y_{\rm cut}$ is some constant.  We
accept events as corresponding to the signal hypothesis if they are on
one side of the boundary, e.g., $y(\bvec{x}) \le y_{\rm cut}$ could
represent the acceptance region and $y(\bvec{x}) > y_{\rm cut}$ could
be the rejection region.

Equivalently we can use the function $y(\bvec{x})$ as a scalar {\it
  test statistic}.  Once its functional form is specified, we can
determine the pdfs of $y(\bvec{x})$ under both the signal and
background hypotheses, $p(y|s)$ and $p(y|b)$.  The
decision boundary is now effectively a single cut on the scalar
variable $y$, as illustrated in Fig.~\ref{fig:TestStat}.

\setlength{\unitlength}{1.0 cm}
\renewcommand{\baselinestretch}{0.8}
\begin{figure}[htbp]
\begin{picture}(10.0,5.5)
\put(0,-0.2){\includegraphics{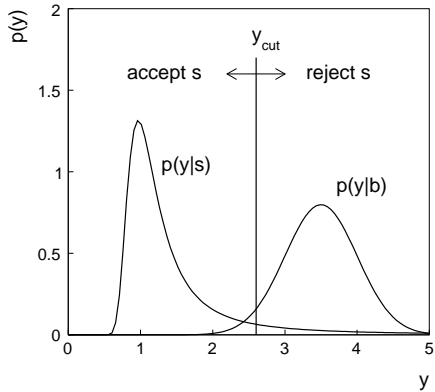}}
\put(6.7,0.2){\makebox(5,4)[b]{\begin{minipage}[b]{5cm}
\protect\caption{{\small Distributions of the scalar
test statistic $y(\bvec{x})$ under the signal and background
hypotheses.}
\protect\label{fig:TestStat}}
\end{minipage}}}
\end{picture}
\end{figure}
\renewcommand{\baselinestretch}{1}
\small\normalsize

We would like to design a test to have a high probability to reject a
hypothesis if it is false, which is what we have called the power of
the test.  Unfortunately a test with maximum power with respect to one
alternative will not be optimal with respect to others, so there is no
such thing as an ideal ``model-independent'' test.  Nevertheless, for
a specific pair of signal and background hypotheses, it turns out that
there is a well defined optimal solution to our problem.  The {\it
  Neyman--Pearson} lemma (see, e.g., Ref.~\cite{bib:KS}) states that
for a test of a given significance level of the background hypothesis
(i.e., fixed background efficiency), one obtains the maximum power
relative to the signal hypothesis (signal efficiency) by defining the
critical region $w$ such that for $\bvec{x} \in w$ the {\it likelihood
  ratio}, i.e., the ratio of pdfs for signal and background,

\begin{equation}
\label{eq:lratio}
y(\bvec{x} ) = 
\frac{f(\bvec{x} | s )}{ f(\bvec{x} |b )} \;,
\end{equation}

\noindent is greater than or equal to a given constant, and it is less
than this constant everywhere outside the critical region.  This is
equivalent to the statement that the ratio~(\ref{eq:lratio})
represents the test statistic with which one obtains the highest
signal efficiency for a given background efficiency.

In principle the signal and background theories should allow us to
work out the required functions $f(\bvec{x} | s)$ and $f(\bvec{x} |
b)$, but in practice the calculations are too difficult and we do not
have explicit formulae for these.  What we have instead of $f(\bvec{x}
| s)$ and $f(\bvec{x} | b)$ are complicated Monte Carlo programs, from
which we can sample $\bvec{x}$ to produce simulated signal and
background events.  Because of the multivariate nature of the data,
where $\bvec{x}$ may contain at least several or perhaps even hundreds
of components, it is a nontrivial problem to construct a test with a
power approaching that of the likelihood ratio.

In the usual case where the likelihood ratio~(\ref{eq:lratio}) cannot
be used explicitly, there exist a variety of other multivariate
classifiers such as neural networks, boosted decision trees and
support vector machines that effectively separate different types of
events.  Descriptions of these methods can be found, for example, in
the textbooks~\cite{bib:Bishop,bib:Hastie,bib:Duda,bib:Webb}, lecture
notes~\cite{bib:Cowan09} and proceedings of the PHYSTAT conference
series~\cite{bib:PHYSTAT}.  Software for HEP includes the {\tt
  TMVA}~\cite{bib:TMVA} and {\tt
  StatPatternRecognition}~\cite{bib:Narsky05} packages.

\section{Frequentist treatment of discovery and limits}
\label{sec:disclim}

The use of a statistical test to in a Particle Physics analysis
involving different event types comes up in different ways.  Sometimes
both event classes are known to exist, and the goal is to select one
class (signal) for further study.  For example, top-quark production
in proton--proton collisions is a well-established process.  By
selecting these events one can carry out precise measurements of the
top quark's properties such as its mass.  This was the basic picture
in the previous section.  The measured quantities referred to
individual events, and we tested the the hypothesized event type for
each.

In other cases, the signal process could represent an extension to the
Standard Model, say, supersymmetry, whose existence is not yet
established, and the goal of the analysis is to see if one can do
this.  Here we will imagine the ``data'' as representing not
individual events but a sample of events, i.e., an entire
``experiment''.  If the signal process we are searching for does not
exist, then our sample will consist entirely of background events,
e.g., those due to Standard Model processes.  If the signal does
exist, then we will find both signal and background events.  Thus the
hypothesis we want to test is

\[
H_0 \; : \; \mbox{only background processes exist}
\]

\noindent versus the alternative

\[
H_1 \; : \; \mbox{both signal and background exist} .
\]

We will refer to the hypothesis $H_0$ as the background-only model (or
simply ``$b$'') and the alternative $H_1$ as the signal-plus-background
model, $s+b$.  The Neyman-Pearson lemma still applies.  In a test of
$H_0$ of a given size, the highest power relative to $H_1$ is obtained
when the critical region contains the highest values of the likelihood
ratio $L(H_1)/L(H_0)$.  Here, however, the likelihood is the
probability for the entire set of data from the experiment, not just
for individual events.

Rejecting $H_0$ means in effect discovering a new phenomenon.  Of
course before we believe that we have made a new discovery, a number
of other concerns must be addressed, such as our confidence in the
reliability of the statistical models used, the plausibility of the
new phenomenon and the degree to which it can describe the data.  Here
however we will simply focus on question of statistical significance
and in effect equate ``rejecting the background-only hypothesis'' with
``discovery''.  Often in HEP one claims discovery when the $p$-value
of the background-only hypothesis is found below $2.9 \times 10^{-7}$,
corresponding to a 5-sigma effect.  We will revisit the rationale
behind this threshold in Sec.~\ref{sec:why5}.

Even if one fails to discover new physics by rejecting the
background-only model, one can nevertheless test various signal models
and see whether they are compatible with the data.  Signal models are
usually characterized by some continuous parameters representing,
e.g., the masses of new particles.  If we carry out a test of size
$\alpha$ for all possible values of the parameters, then those that
are not rejected constitute what is called a {\it confidence region}
for the parameters with a {\it confidence level} of $\mbox{CL} = 1 -
\alpha$.  By construction a hypothesized point in parameter space
will, if it is true, be rejected with probability $\alpha$.  Therefore
the confidence region will contain the true value of the parameters
with probability $1 - \alpha$.  For purposes of confidence limits one
typically uses a test of size $\alpha = 0.05$, which is to say the
regions have a confidence level of 95\%.

If the problem has only one parameter, then the region is called a
confidence interval.  An important example is where a parameter $\mu$
is proportional to the cross section for the signal process whose
existence is not yet established.  Here one often wants to test a
hypothetical value relative to the alternative hypothesis that the
signal does not exist, i.e., $\mu = 0$.  The critical region of the
test is then taken have higher probability for the lower values of the
parameter.

For example, suppose the data consist of a value $x$ that follows a
Gaussian distribution with unknown mean $\mu$ and known standard
deviation $\sigma$.  If we test a value $\mu$ relative to the
alternative of a smaller value, then the critical region will consist
of values of $x < c$ for some constant $c$ such that

\begin{equation}
\label{eq:gausscrit}
\alpha = \int_{-\infty}^{c} \frac{1}{\sqrt{2 \pi} \sigma} 
e^{- (x - \mu)^2 / 2 \sigma^2} \, dx
= \Phi \left( \frac{ c - \mu }{\sigma} \right) \;,
\end{equation}

\noindent or

\begin{equation}
\label{eq:gausscrit2}
c = \mu - \sigma \Phi^{-1}(1 - \alpha) \;.
\end{equation}

\noindent If we take, e.g., $\alpha = 0.05$, then the factor
$\Phi^{-1}(1 - \alpha) = 1.64$ says that the critical region starts at
1.64 standard deviations below the value of $\mu$ being tested.  If
$x$ is observed any lower than this, then the corresponding $\mu$ is
rejected.

Equivalently we can take the $p$-value of a hypothesized $\mu$,
$p_{\mu}$, as the probability to observe $x$ as low as we found or
lower, and we then reject $\mu$ if we find $p_{\mu} \le \alpha$.  The
highest value of $\mu$ that we do not reject is called the {\it upper
  limit} of $\mu$ at a confidence level of $1 - \alpha$, and we will
write this here as $\mu_{\rm up}$ Lower limits $\mu_{\rm lo}$ can of
course be constructed using an analogous procedure.  In practice these
points are found by setting $p_{\mu} = \alpha$ and solving for $\mu$.
There are a number of subtle issues connected with limits derived in
this way and we will return to these in Sec.~\ref{sec:spurious}.

\subsection{A toy example}

Consider an experiment where we measure for each selected event two
quantities, which we can write as a vector $\bvec{x} = (x_1, x_2)$.
Suppose that for background events $\bvec{x}$ follows

\begin{equation}
\label{eq:fxb}
f(\bvec{x} | b) = \frac{1}{\xi_1} e^{-x/\xi_1}
\frac{1}{\xi_2} e^{- x/\xi_2} \;,
\end{equation}

\noindent and for a certain signal model they follow

\begin{equation}
\label{eq:fxs}
f(\bvec{x} | s) = C \frac{1}{\sqrt{2 \pi} \sigma_1}
e^{-(x_1 - \mu_1)^2 / 2 \sigma_1^2} \,
\frac{1}{\sqrt{2 \pi} \sigma_2}
e^{-(x_2 - \mu_2)^2 / 2 \sigma_2^2} \;,
\end{equation}

\noindent where $x_1 \ge 0$, $x_2 \ge 0$ and $C$ is a
normalization constant.  The distribution of events generated 
according to these hypotheses are shown in Fig.~\ref{fig:fxsb}(a).

\setlength{\unitlength}{1.0 cm}
\renewcommand{\baselinestretch}{0.9}
\begin{figure}[htbp]
\begin{picture}(10.0,5.5)
\put(-0.5,0){\includegraphics[width=0.45\textwidth]{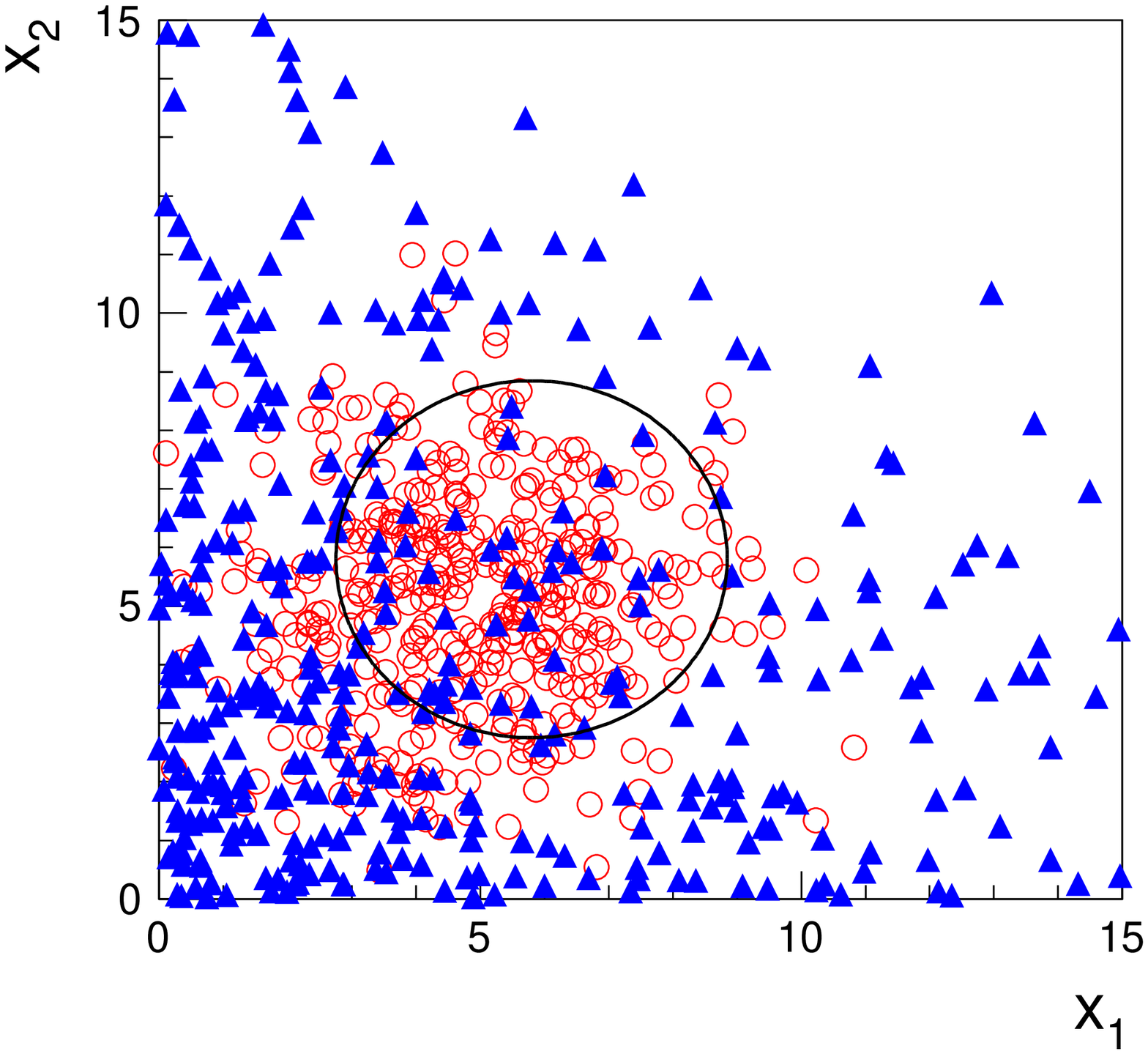}}
\put(6.,0){\includegraphics[width=0.45\textwidth]{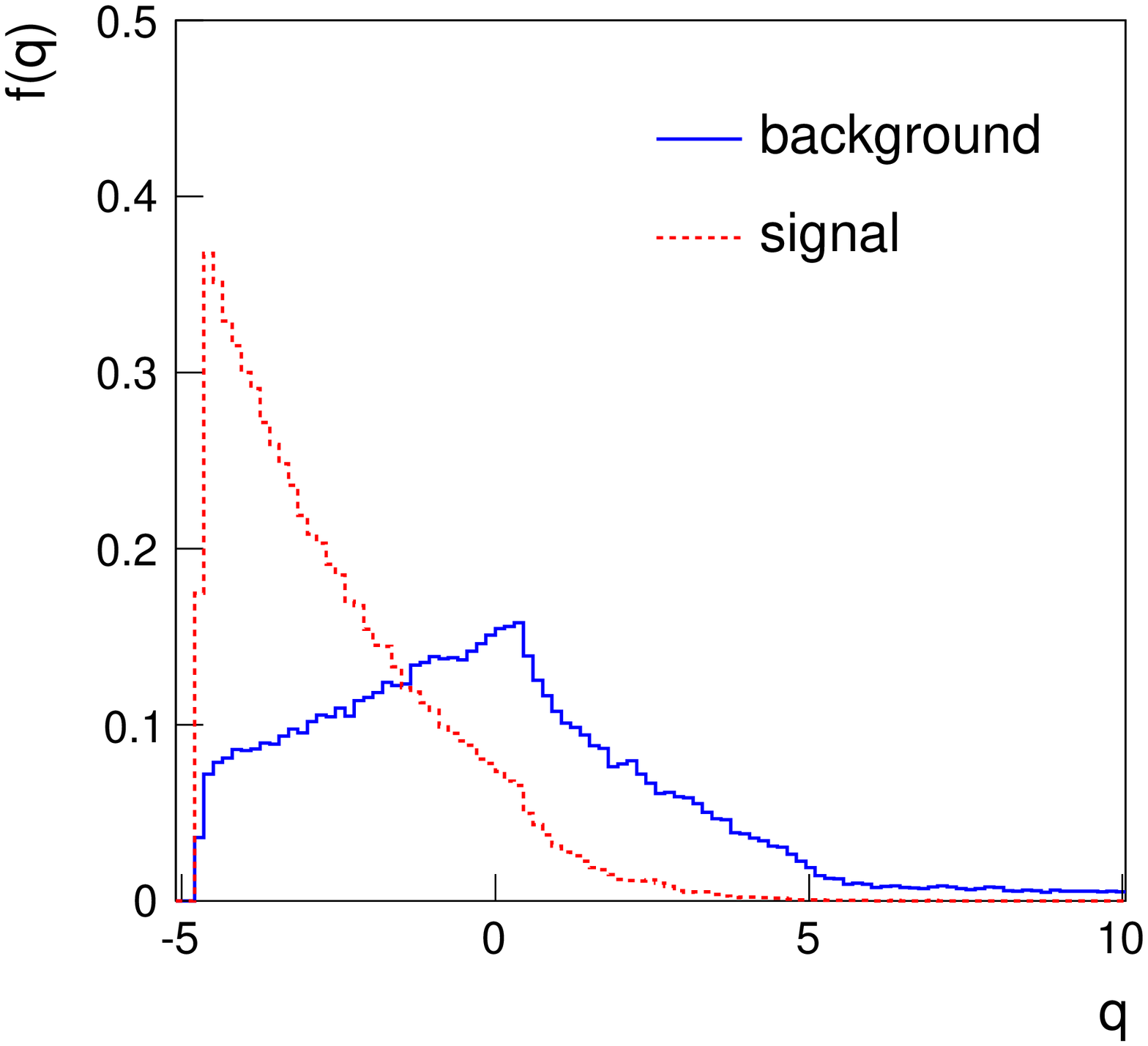}}
\put(4.7,4.5){(a)}
\put(11.3,4.5){(b)}
\end{picture}
\caption{\small (a) Distributions of $\bvec{x} = (x_1, x_2)$ for
  events of type signal (red circles) and background (blue triangles)
  shown with a contour of constant likelihood ratio; (b) the
  distribution of the statistic $q$ for signal and background events.}
\label{fig:fxsb}
\end{figure}
\renewcommand{\baselinestretch}{1}
\small\normalsize

First, suppose that the signal and background both correspond to event
types that are known to exist and the goal is simply to select signal.
In this case we can exploit the Neyman-Pearson lemma and base the
selection on the likelihood ratio

\begin{equation}
y(\bvec{x}) = \frac{f(\bvec{x} | s)}{f(\bvec{x} | b)} \;.
\end{equation}

\noindent We can define the same critical region by using any
monotonic function of the likelihood ratio, and in this case
it is useful to take

\begin{equation}
  q = \left( \frac{ x_1 - \mu_1}{\sigma_1} \right)^2
  + \left( \frac{x_2 - \mu_2}{\sigma_2} \right)^2
  - \frac{2 x_1}{\xi_1} - \frac{2 x_2}{\xi_2} 
= - 2 \ln y(\bvec{x}) + \mbox{const.}
\end{equation}

\noindent Distributions of the statistic $q$ for the background and
signal hypotheses (\ref{eq:fxb}) and (\ref{eq:fxs}) are shown in
Fig.~\ref{fig:fxsb}(b).  This shows that a sample enhanced in signal
events can be selected by selecting events with $q$ less than a given
threshold, say, $q_{\rm cut}$.

Now suppose instead that the signal process is not known to exist and
the goal of the analysis is to search for it.  Suppose that the
expected numbers events are $b$ of background $s$ for a given signal
model.  For now assume that the model's prediction for both of these
quantities can be determined with negligible uncertainty.  The the
actual number of events $n$ that we find can be modeled as a Poisson
distributed quantity whose mean we can write as $\mu s + b$, where
$\mu$ is a parameter that specifies the strength of the signal
process.  That is, the probability to find $n$ events is

\begin{equation}
\label{eq:pnmusb}
P(n | \mu) = \frac{(\mu s + b)^n}{n!} e^{-(\mu s + b)} \;.
\end{equation}

The values of $\bvec{x}$ follow a pdf that is a mixture
of the two contributions from signal and background,

\begin{equation}
\label{eq:fxsb}
f(\bvec{x} | \mu ) = \frac{\mu s}{\mu s + b} f(\bvec{x} | s) + 
\frac{b}{\mu s + b} f(\bvec{x} |b) \;,
\end{equation}

\noindent where the coefficients of each component give the fraction of
events of each type.

The complete measurement thus consists of selecting $n$ events and for
each one measuring the two-dimensional vector quantity $\bvec{x}$.
The full likelihood is therefore

\begin{equation}
\label{eq:fullL}
L(\mu) =  P(n | \mu) \prod_{i=1}^{n} f(\bvec{x}_i | \mu) = 
\frac{e^{-(\mu s+b)}}{n!}  \prod_{i=1}^n
\left[ \mu s f(\bvec{x}_i | s) + bf(\bvec{x}_i | b) \right] \;.
\end{equation}

We can now carry out tests of different hypothetical values of $\mu$.
To establish the existence of the signal process we try to reject the
hypothesis of the background-only model, $\mu=0$.  Regardless of
whether we claim discovery we can set limits on the signal strength
$\mu$, which we examine further in Sec.~\ref{sec:spurious}.

Let us first focus on the question of discovery, i.e., a test of
$\mu=0$.  If the signal process exists, we would like to maximize the
probability that we will discover it.  This means that the test of the
background-only ($\mu=0$) hypothesis should have as high a power as
possible relative to the alternative that includes signal ($\mu=1$).
According to the Neyman-Pearson lemma, the maximum power is achieved
by basing the test on the likelihood ratio $L(1)/L(0)$, or
equivalently on the statistic

\begin{equation}
\label{eq:Qsbdef}
Q = - 2 \ln \frac{L(1)}{L(0)} = 
-s + \sum_{i=1}^n \ln \left( 1 + \frac{s}{b} 
\frac{f(\bvec{x}_i | s)}{f(\bvec{x}_i | b)} \right) \;.
\end{equation}

\noindent The term $-s$ in front of the sum is a constant and so only
shifts the distribution of $Q$ for both hypotheses equally; it can
therefore be dropped.

The other terms on the right-hand side of Eq.~(\ref{eq:Qsbdef}) are a
sum of contributions from each event, and because the $\bvec{x}$
values follow the same distribution for each event, each term in the
sum follows the same distribution.  To find the pdf of $Q$ we can
exploit the fact that the distribution of a sum of random variables is
given by the convolution of their distributions.  The full
distribution can therefore be determined using Fourier transform
techniques from the corresponding single-event distributions; details
can be found in Ref.~\cite{bib:HuNielsen}.

Following our toy example, suppose we take the expected numbers of
events to be $b = 100$ for background and $s = 20$ for signal.  The
distribution of the statistic $Q$ is found in this case simply by
generating experiments according to the $\mu=0$ and $\mu=1$
hypotheses, computing $Q$ for each according to Eq.~(\ref{eq:Qsbdef})
and recording the values in histograms.  This results in the
distributions shown in Fig.~\ref{fig:Qdist3}(a).

\setlength{\unitlength}{1.0 cm}
\renewcommand{\baselinestretch}{0.9}
\begin{figure}[htbp]
\begin{picture}(10.0,5.25)
\put(0.,0){\includegraphics[width=0.45\textwidth]{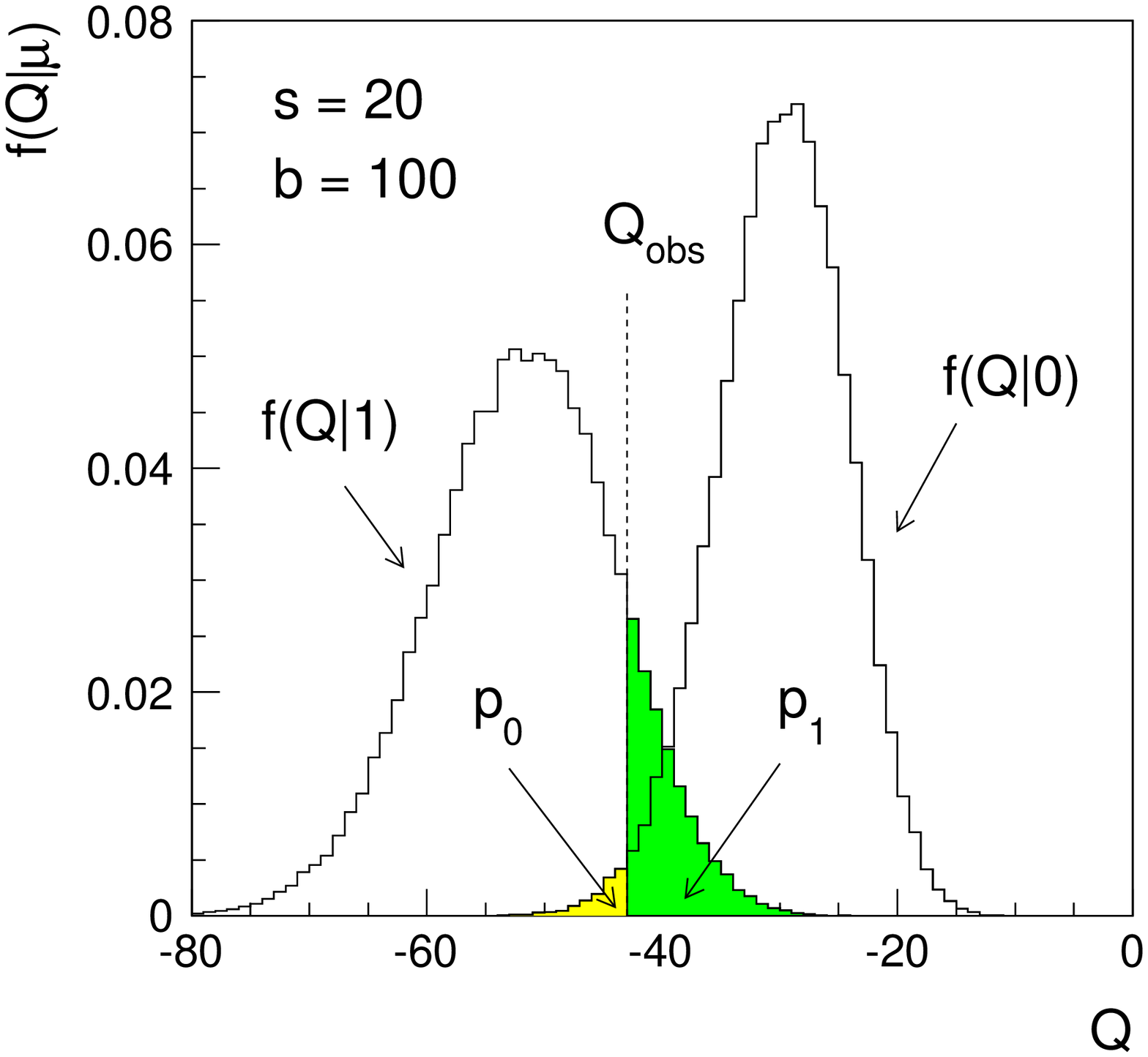}}
\put(6.,0){\includegraphics[width=0.45\textwidth]{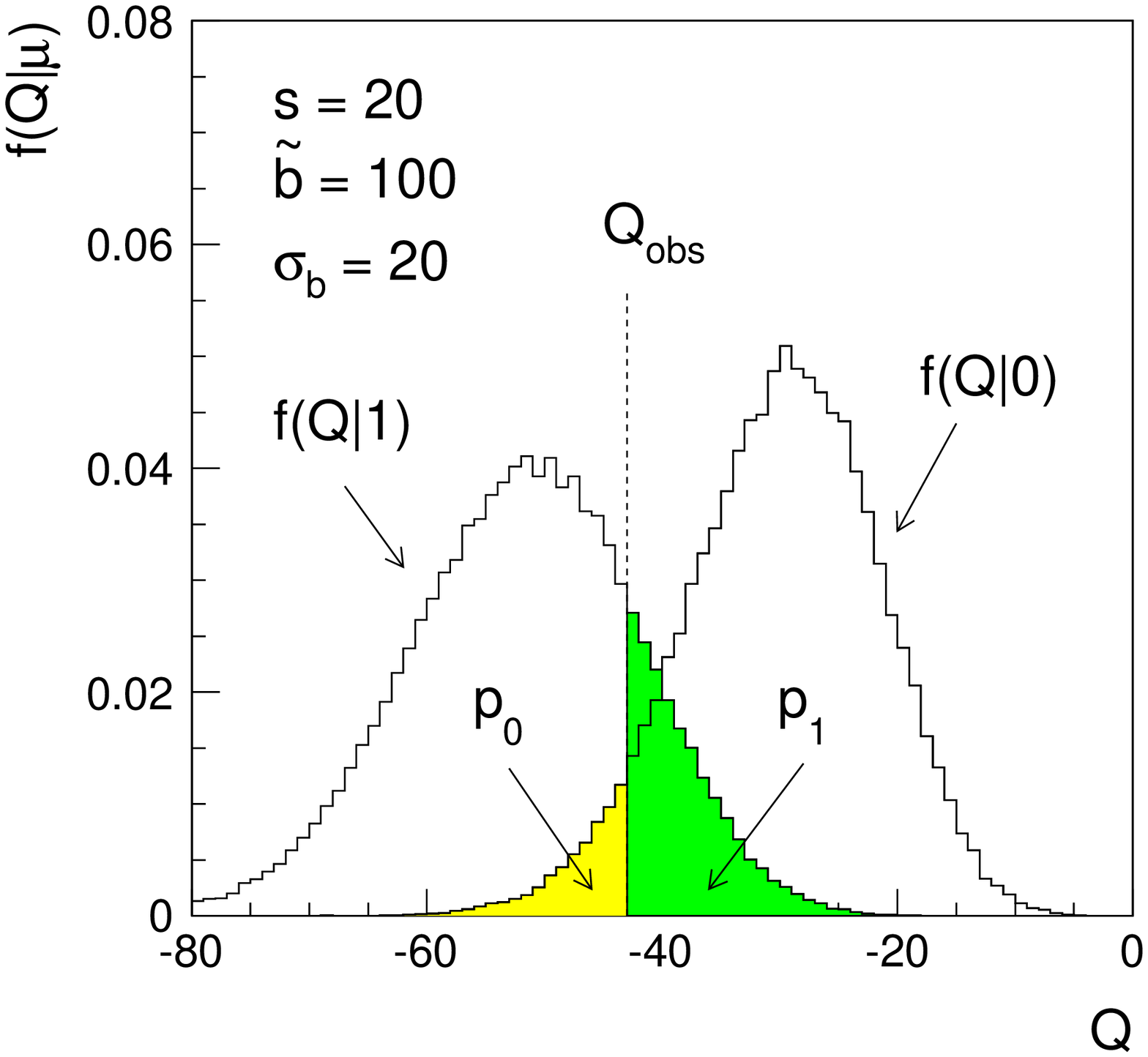}}
\put(5.2,4.5){(a)}
\put(11.2,4.5){(b)}
\end{picture}
\caption{\small (a) Distribution of the statistic $Q$ assuming $s =
  20$ and $b = 100$ under both the background-only ($\mu=0$)
  signal-plus-background ($\mu=1$) hypotheses; (b) same as in (a) but
  with $b$ treated as having an uncertainty of $\sigma_b = 20$ (see
  text).}
\label{fig:Qdist3}
\end{figure}
\renewcommand{\baselinestretch}{1}
\small\normalsize

To establish discovery of the signal process, we use the statistic $Q$
to test the hypothesis that $\mu=0$.  As the test statistic is a
monotonic function of the likelihood ratio $L(1)/L(0)$, we obtain
maximum power relative to the alternative of $\mu=1$.  The $p$-vale of
$\mu=0$ is computed as the area below $Q_{\rm obs}$ in
Fig.~\ref{fig:Qdist3}, i.e. $p_0 = P(Q \le Q_{\rm obs} | 0)$, because
here lower $Q$ corresponds to data more consistent with a positive
$\mu$ (e.g., $\mu=1$).  The $p$-value can be converted into a
significance $Z$ using Eq~(\ref{eq:significance}) and if this is
greater than a specific threshold (e.g., 5.0) then one rejects the
background-only hypothesis.

To set limits on $\mu$ we can use the statistic

\begin{equation}
Q_{\mu} = - 2 \ln \frac{L(\mu)}{L(0)} \;,
\end{equation}

\noindent defined such that the special case $Q_1$ is the same as the
statistic $Q$ used above for discovery.  This will provide maximum
power in a test of $\mu$ relative to the background-only alternative.
The distribution of $Q$ is also shown in Fig.~\ref{fig:Qdist3} for
$\mu = 1$.  The $p$-value of the $\mu=1$ hypothesis is given by the
area above the observed value $Q_{\rm obs}$, since higher values of
$Q$ are more consistent with the alternative of $\mu=0$.  This is
shown here for the special case of $\mu=1$ but one can repeat the
procedure using $f(Q_{\mu}|\mu)$ for any other value of $\mu$ and
compute the $p$-value $p_{\mu}$ in the analogous manner.  To find the
upper limit one would carry out the analysis as described above for
all values of $\mu$ and reject those that have $p_{\mu} < \alpha$ for,
say $\alpha = 0.05$.  The highest value of $\mu$ not rejected is then
the upper limit at 95\% C.L.


\subsection{Systematic uncertainties and nuisance parameters}
\label{sec:sysnuis}

Until now we have treated the expected number of background events $b$
as known with negligible uncertainty.  In practice, of course, this
may not be true and so we may need to regard $b$ as an adjustable
parameter of our model.  That is, we regard Eq.~(\ref{eq:fullL}) as
giving $L(\mu, b)$, where $\mu$ is the parameter of interest and $b$
is a {\it nuisance parameter}.

There are several ways of eliminating the nuisance parameters from the
problem.  First we consider a method that is essentially frequentist
but contains a Bayesian aspect.  From the standpoint of nuisance
parameters this is essentially the same as what is done in the purely
Bayesian limits as discussed in Sec.~\ref{sec:bayeslim}.  An
alternative frequentist treatment using the profile likelihood is
described in Sec.~\ref{sec:pl}.  

Consider first the frequentist method with Bayesian treatment of
nuisance parameters.  Our degree of belief about the true value of the
parameter may be described in a Bayesian sense by a prior pdf $\pi(b)$
and our best estimate of $b$ (e.g., the mean of $\pi(b)$) may be a
value $\tilde{b}$.  As an example, $\pi(b)$ could be a Gaussian
distribution centred about $\tilde{b}$ with a standard deviation
$\sigma_{b}$:

\begin{equation}
\label{eq:pib}
  \pi(b) = \frac{1}{\sqrt{2 \pi} \sigma_b} 
  e^{-(b - \tilde{b})^2 / 2 \sigma_b^2} \;.
\end{equation}

\noindent In fact a Gaussian pdf for $b$ may not be the most
appropriate model, e.g., if a parameter is bounded to be positive or
if the prior should be characterized by longer positive tail.  As an
alternative one may use a Gaussian distribution for $\ln b$, which is
to say that the pdf for $b$ is log-normal.

Using the pdf $\pi(b)$ we can construct what is called the marginal
(or prior predictive) likelihood,

\begin{equation}
\label{eq:Lmmu}
L_{\rm m}(n, \bvec{x}_1, \ldots, \bvec{x}_n | \mu) = 
\int L(n, \bvec{x}_1, \ldots, \bvec{x}_n | \mu, b)  \pi(b) \, db \;,
\end{equation}

\noindent where in the notation above we have emphasized that the
likelihood of a model is the probability for the data under assumption
of that model.  

Notice that the marginal model does not represent the probability of
data that would be generated if we were really to repeat the
experiment.  In that case we would not know the true value of $b$, but
we could at least assume it would not change under repetition of the
experiment.  Rather, the marginal model represents a situation in
which every repetition of the experiment is carried out with a new
value of $b$ randomly sampled from $\pi(b)$.  It is in effect an
average of models each with a given $b$, where the average is carried
out with respect to the density $\pi(b)$.

For our tests we can use the same test statistic $Q$ as before, but
now we need to know its distribution under assumption of the prior
predictive model.  That is, if $b$ is known exactly then we obtain
distributions $f(Q | \mu, b)$ such as those shown in
Fig.~\ref{fig:Qdist3}(a).  What we want instead is the distribution
based on data that follows the marginal model,

\begin{equation}
\label{eq:fmQ}
f_{\rm m} (Q | \mu) = \int f(Q |\mu, b) \pi(b) \, db \;.
\end{equation}

\noindent Although it may not be obvious how to compute this integral,
it can be done easily with Monte Carlo by generating a value of $b$
according to $\pi(b)$, then using this value to generate the data $n$,
$\bvec{x}_1, \ldots, \bvec{x}_n$, and with these we find a value of
$Q$ which is recorded in a histogram.  By repeating the entire
procedure a large number of times we obtain distributions as shown in
Fig.~\ref{fig:Qdist3}(b), which are generated with a Gaussian prior
for $b$ with $\tilde{b} = 100$ and $\sigma_b =20$.

As can be seen in Fig.~\ref{fig:Qdist3}, the effect of the uncertainty
on $b$ broadens the distributions of $Q$ such that the $p$-values for
both hypotheses are increased.  That is, one may be able to reject one
or the other hypothesis in the case where $b$ was known because the
$p$-value may be found less than $\alpha$.  When the uncertainty in
$b$ is included, however, the $p$-values may no longer allow one to
reject the model in question.

As a further step one could consider using the marginal likelihood as
the basis of the likelihood ratio used in the test statistic, i.e., we
take $Q = -2 \ln (L_{\rm m}(1) / L_{\rm m}(0))$.  Use of a different
statistic simply changes the critical region of the test and thus
alters the power relative to the alternative models considered.  This
step by itself, however, does not take into account the uncertainty in
$b$ and it will not result in a broadening of $f(Q|\mu)$ and an
increase in $p$-values as illustrated above.  This is achieved by
generating the distribution of $Q$ using the marginal model through
Eq.~(\ref{eq:fmQ}).  In practice the marginal likelihoods can be very
difficult to compute and a test statistic based on their ratio is not
often used in HEP (see, however, Ref.~\cite{bib:severini}).

The ratio of marginal likelihoods is also called the {\it Bayes
  factor}, usually written with indices to denote the hypotheses being
compared, e.g.,

\begin{equation}
\label{eq:bayesfactor}
B_{10} = \frac{L_{\rm m}(1)}{L_{\rm m}(0)} \;.
\end{equation}

\noindent This is by itself a quantity of interest in Bayesian
statistics as it represents the ratio of posterior probabilities of
the hypotheses $\mu=1$ and $\mu=0$ in the special case where the prior
probabilities are taken equal.  If the Bayes factor is greater than
one it means that the evidence from the data results in an increase in
one's belief in the hypothesis $\mu=1$ over $\mu=0$.  Further
discussion on the use of Bayes factors can be found in
Refs.~\cite{bib:PDG,bib:KassRaftery}.

Another possibility is to construct the test statistic from the ratio
of {\it profile likelihoods}.  Suppose the likelihood depends on a
parameter of interest $\mu$ and nuisance parameters $\bvec{\theta} =
(\theta_1, \ldots, \theta_N)$.  The profile likelihood $L_{\rm p}$ is
defined as

\begin{equation}
\label{eq:pl}
L_{\rm p}(\mu) = L(\mu, \hat{\hat{\bvec{\theta}}}(\mu)) \;,
\end{equation}

\noindent where $\hat{\hat{\bvec{\theta}}}(\mu)$, called the profiled
values of the nuisance parameters $\bvec{\theta}$, are the values that
maximizes $L(\mu, \theta)$ for the specified value of $\mu$.  Thus the
profile likelihood only depends on $\mu$.  Searches at the Tevatron
(e.g., Ref.~\cite{bib:qtev}) have used the statistic

\begin{equation}
Q = - 2 \ln \frac{L_{\rm p}(1)}{L_{\rm p}(0)} \;.
\end{equation}

\noindent As mentioned above, use of this statistic does not in itself
take into account the systematic uncertainties related to the nuisance
parameters.  In Ref.~\cite{bib:qtev} this has been done by generating
the distribution of $Q$ using the marginal model (\ref{eq:Lmmu}).  An
alternative to this procedure is to construct the statistic from a
different profile likelihood ratio as described in Sec.~\ref{sec:pl}.

\section{Tests based on the profile likelihood ratio}
\label{sec:pl}

Suppose as before that the parameter of interest is $\mu$ and the
problem may contain one or more nuisance parameters $\bvec{\theta}$.
An alternative way to test hypothetical values of $\mu$ is to use
the {\it profile likelihood ratio},

\begin{equation}
\label{eq:plr}
\lambda(\mu) = \frac{L_{\rm p}(\mu)}
{L(\hat{\mu}, \hat{\theta})} \;,
\end{equation}

\noindent where $L_{\rm p}$ is the profile likelihood defined in
Eq.~(\ref{eq:pl}) and $\hat{\mu}$ and $\hat{\theta}$ are the values of
the parameters that maximize the likelihood.  In some models it may be
that $\mu$ can only take on values in a restricted range, e.g., $\mu
\ge 0$ if this parameter is proportional to the cross section of the
signal process.  In this case we can, however, regard $\hat{\mu}$ as
an effective estimator that is allowed to take on negative values.
This will allow us to write down simple formulae for the distributions
of test statistics that are valid in the limit where the data sample
is very large.

The quantity $\lambda(\mu)$ is defined so that it lies between zero
and one, with higher values indicating greater compatibility between
the data and the hypothesized value of $\mu$.  We can therefore use
$\lambda(\mu)$ to construct a statistic to test different values of
$\mu$.  Suppose as above that $\mu$ is proportional to the rate of the
sought after signal process and we want to test the background-only
($\mu=0$) hypothesis.

Often the signal process is such that only positive values of $\mu$
are regarded as relevant alternatives.  In this case we would choose
the critical region of our test of $\mu = 0$ to correspond to data
outcomes characteristic of positive $\mu$, that is, when $\hat{\mu} >
0$.  It could happen that we find $\hat{\mu} < 0$, e.g., if the total
observed number of events fluctuates below what is expected from
background alone.  Although a negative $\hat{\mu}$ indicates a level
of incompatibility between the data and hypothesis of $\mu=0$, this is
not the type of disagreement that we want to exploit to declare
discovery of a positive signal process.

Providing our signal models are of the type described above, we can
take the statistic used to test $\mu=0$ as

\begin{equation}
\label{eq:q0} 
q_{0} =  
\left\{ \! \! \begin{array}{ll}
               - 2 \ln \lambda(0)  
               & \quad \hat{\mu} \ge 0 \;, \\*[0.3 cm]
               0 & \quad \hat{\mu} < 0  \;,
              \end{array}
       \right.
\end{equation}

\noindent where $\lambda(0)$ is the profile likelihood ratio for
$\mu=0$ as defined in Eq.~(\ref{eq:plr}).  In this way, higher values
of $q_0$ correspond to increasing disagreement between data and
hypothesis, and so the $p$-value of $\mu=0$ is the probability,
assuming $\mu=0$ to find $q_0$ at least high or higher than the
observed value.

If we are interested in an upper limit for the parameter $\mu$, then
we want the critical region to correspond to data values
characteristic of the alternative $\mu=0$.  This can be achieved by
defining

\begin{equation}
\label{eq:qmu} 
q_{\mu} =  
\left\{ \! \! \begin{array}{ll}
               - 2 \ln \lambda(\mu)  
               & \quad \hat{\mu} \le \mu \;, \\*[0.3 cm]
               0 & \quad \hat{\mu} > \mu  \;.
              \end{array}
       \right.
\end{equation}

\noindent For both discovery and upper limits, therefore, the
$p$-value for a hypothesized $\mu$ is then

\begin{equation}
\label{eq:pmu}
  p_{\mu} = \int_{q_{\mu,{\rm obs}}}^{\infty} f(q_{\mu} | \mu, \theta ) 
  \, dq_{\mu} \;,
\end{equation}

\noindent If we use the statistic $q_{\mu}$ then we find the upper
limit $\mu_{\rm up}$ at confidence level $1 - \alpha$ by setting
$p_{\mu} = \alpha$ and solving for $\mu$.  This will have the property
$P(\mu_{\rm up} \ge \mu) \ge \alpha$.  Note that the $p$-value
pertains to the hypothesis of not only of $\mu$ but also the nuisance
parameters $\vec{\theta}$.  We will return to this point below.

To find the $p$-value we need the distribution of the test statistic
under assumption of the same $\mu$ being tested.  For sufficiently
large data samples one can show that this distributions approaches an
asymptotic form related to the chi-square distribution, where the
number of degrees of freedom is equal to the number of parameters of
interest (in this example just one, i.e., $\mu$).  The asymptotic
formulae are based on theorems due to Wilks \cite{bib:wilks} and Wald
\cite{bib:wald} and are described in further detail in
Ref.~\cite{bib:asimov}.

An important advantage of using the profile likelihood ratio is that
its asymptotic distribution is independent of the nuisance parameters,
so we are not required to choose specific values for them to compute
the $p$-value.  In practice one has of course a finite data sample and
so the asymptotic formulae are not exact.  Therefore the $p$-values
will in general depend on the nuisance parameters to some extent.

Providing the conditions for the asymptotic approximations hold, one
finds a very simple formula for the $p$-value,

\begin{equation}
\label{eq:pmua}
p_{\mu} = \Phi \left( \sqrt{ q_{\mu}} \right) \;,
\end{equation}

\noindent where $\Phi$ is the cumulative distribution of the standard
Gaussian.  From Eq.~(\ref{eq:significance}) we find for the
corresponding significance

\begin{equation}
\label{eq:Zmua}
  Z_{\mu} = \sqrt{q_{\mu}} \;.
\end{equation}

For discovery, we could require $Z_0$ greater than some threshold such
as 5.0, which corresponds to $p_0 < 2.9 \times 10^{-7}$.  When setting
limits one usually excludes a parameter value if its $p$-value is less
than, say, 0.05, corresponding to a confidence level of 95\%, or a
significance of 1.64.  Although Eqs.~(\ref{eq:pmua}) and
(\ref{eq:Zmua}) are only exact for an infinitely large data sample,
the approach to the asymptotic limit is very fast and the
approximations often turn out to be valid for moderate or even
surprisingly small data samples.  Examples can be found in
Ref.~\cite{bib:asimov}.

For data samples not large enough to allow use of the asymptotic
formulae, one must determine the distribution of the test statistics
by other means, e.g., with Monte Carlo models that use specific values
for the nuisance parameters.  In the exact frequentist approach we
would then only reject a value of $\mu$ if we find its $p$-value less
than $\alpha$ for all possible value of the nuisance parameters.
Therefore only a smaller set of $\mu$ values are rejected and the
resulting confidence interval becomes larger, which is to say the
limits on $\mu$ become less stringent.  The confidence interval then
{\it overcovers}, i.e., its probability to contain the true $\mu$ is
greater than $1 - \alpha$, at least for some values of the nuisance
parameters.

It may seem unfortunate if we cannot reject values of $\mu$ that are
retained only under assumption of nuisance parameter values that may
be highly disfavoured, e.g., for theoretical reasons.  A compromise
solution is test $\mu$ using the $p$-value based only on the profiled
values of the nuisance parameters, i.e., we take

\begin{equation}
\label{eq:pmupc}
p_{\mu} = \int_{q_{\mu,{\rm obs}}}^{\infty} f(q_{\mu} | \mu, 
\hat{\hat{\theta}}(\mu) ) 
\, dq_{\mu} \;.
\end{equation}

\noindent This procedure has been called {\it profile construction}
\cite{bib:cranmerPC} in HEP or {\it hybrid resampling}
\cite{bib:ChuangLai,bib:Sen} amongst statisticians.  If the true
values of of the nuisance parameters are equal to the profiled values,
then the coverage probability of the resulting confidence interval for
$\mu$ is exact.  For other values of $\theta$ the interval for $\mu$
may over- or undercover.  In cases where it this is crucial one may
include a wider range of nuisance parameter values and study the
coverage with Monte Carlo.

\section{Summary on likelihood ratios}
\label{sec:lrsummary}

Above we have seen two closely related ways to construct tests based
on likelihood ratios and also two different ways of incorporating
systematic uncertainties.  In Sec.~\ref{sec:disclim} we used the ratio
of two simple hypotheses, namely $L(\mu) / L(0)$, whereas in
Sec.~\ref{sec:pl} the statistic used was $L(\mu)/L(\hat{\mu})$.  

If there are no nuisance parameters in the model, then the
Neyman-Pearson lemma guarantees that the ratio $L(\mu) / L(0)$
provides the greatest power in a test of $\mu=0$ with respect to the
alternative of $\mu$.  If there are nuisance parameters then this will
not in general hold.  In this case one can replace the likelihood $L$
by the marginal or profile likelihood, which results in a different
critical region for the test.  It can be difficult find the exact
power for different alternatives but one can study this using Monte
Carlo.  The important point is that by changing the critical region of
the test by using a ratio of marginal or profile likelihoods one does
not by this step alone account for the systematic uncertainties.

To include the uncertainties reflected by the nuisance parameters into
the test we have also seen two approaches.  One has been to construct
the marginal (prior predictive) model (\ref{eq:Lmmu}) to determine the
distribution of the test statistic.  If, for example, one rejects the
hypothesis $\mu=0$, then the model rejected represents an average of
models corresponding to different values of the nuisance parameters.
This is the approach we used together with the likelihood ratio
$L(\mu)/L(0)$ (with or without the marginal or profile likelihoods in
the ratio).

In contrast to this, when we used the statistic based on the profile
likelihood ratio $L(\mu, \hat{\hat{\bvec{\theta}}}) / L(\hat{\mu},
\hat{\bvec{\theta}})$ we exploited the fact that its distribution
becomes independent of the nuisance parameters in the large sample
limit.  In this case we are able to say that if $\mu=0$ is rejected,
then this holds for all values of the nuisance parameters
$\bvec{\theta}$.  The large-sample distributions based on Wilks'
theorem are only valid when the likelihood ratio is constructed in
this way; this is not the case, e.g., if one were to characterize
nuisance parameters with a prior and then marginalize (integrate).

In a real analysis the data sample is finite and so the $p$-values for
the parameter of interest $\mu$ will depend at some level on the
nuisance parameters $\bvec{\theta}$.  In such a case one may then use
their profiled values $\hat{\hat{\bvec{\theta}}}$ under assumption of
the value of $\mu$ being tested.  If these are equal to the true
values of the nuisance parameters, then the $p$-values for $\mu$ will
be correct and a confidence interval for $\mu$ will cover the true
value with a probability equal to the nominal confidence level $1 -
\alpha$.  If the true values of $\bvec{\theta}$ are not equal to the
profiled values, then the $p$-values may be too high or to low, which
is to say that confidence intervals for $\mu$ may be to large or too
small.

Both types of likelihood ratios and more importantly, both methods for
determining their sampling distributions (averaged or not) are widely
used.  For many analyses they will lead to very similar conclusions.
An important advantage of the profile likelihood ratio is that one can
say what set of physical models have been rejected (i.e., what points
in nuisance parameter space).  If necessary, Monte Carlo studies can
be carried out to obtain the $p$-values using nuisance parameters in
some region about their profiled values
$\hat{\hat{\bvec{\theta}}}(\mu)$.

\section{Unified intervals}
\label{sec:fc}

The test of $\mu$ used for an upper limit assumes that the relevant
alternative hypothesis is $\mu=0$, and the critical region is chosen
accordingly.  In other cases one may regard values of $\mu$ both
higher and lower than the one being tested as valid alternatives, and
one would therefore like a test that has high power for both cases.
One can show that in general there is no single test (i.e., no given
critical region) that will have the highest power relative to all
alternatives (see, e.g., Ref.~\cite{bib:KS}, Chapter 22).

Nevertheless we can use the statistic

\begin{equation}
\label{eq:tmu}
t_{\mu} = -2 \ln \lambda(\mu)
\end{equation}

\noindent to construct a test for any value of $\mu$.  As before,
higher values of the statistic correspond to increasing disagreement
between the data and the hypothesized $\mu$.  Here, however, the
critical region can include data corresponding to an estimated signal
strength $\hat{\mu}$ greater or less than $\mu$.  If one carries out a
test of all values of $\mu$ using this statistic, then both high and
low values of $\mu$ may be rejected.

Suppose the lowest and highest values not rejected are $\mu_1$ and
$\mu_2$, respectively.  One may be tempted to interpret the upper edge
of such an interval as an upper limit in the same sense as the one
derived above using $q_{\mu}$ from Eq.~(\ref{eq:qmu}).  The coverage
probability, however, refers to the whole interval, i.e., one has
$P(\mu_1 \le \mu \le \mu_2) \ge 1 - \alpha$.  One cannot in general
make a corresponding statement about the probability for the upper or
lower edge of the interval alone to be above or below $\mu$, analogous
to the statement $P(\mu_{\rm up} \ge \mu) \ge 1 - \alpha$ that holds
for an upper limit.

The confidence intervals proposed by Feldman and Cousins
\cite{bib:FC}, also called {\it unified intervals}, are based on a
statistic similar to $t_{\mu}$ from Eq.~(\ref{eq:tmu}) with the
additional restriction that the estimator $\hat{\mu}$ that appears in
the denominator of the likelihood ratio is restricted to physically
allowed values of $\mu$.  Large-sample formulae for the distributions
and corresponding $p$-values can be found in Ref.~\cite{bib:asimov}.
(In that reference the statistic for the case $\mu \ge 0$ is called
$\tilde{t}_{\mu}$.)  The problem of excluding parameter values for
which one has no sensitivity is mitigated with unified intervals by
the particular choice of the critical region of the test (see
Ref.~\cite{bib:FC}).

\section{Bayesian limits}
\label{sec:bayeslim}

Although these lectures focus mainly on frequentist statistical
procedures we provide here a brief description of the Bayesian
approach to setting limits.  This is in fact conceptually much simpler
than the frequentist procedure.  Suppose we have a model that contains
a parameter $\mu$, which as before we imagine as being proportional to
the rate of a sought-after signal process.  In addition the model may
contain some nuisance parameters $\theta$.  As in the frequentist
case, we will have a likelihood $L(\bvec{x} | \mu, \theta)$ which
gives the probability for the data $\bvec{x}$ given $\mu$ and
$\theta$.  In a Bayesian analysis we are allowed to associate a
probability with parameter values, and so we assess our degree of
belief in a given model (or set of parameter values) by giving the
posterior probability $p(\mu, \theta | \bvec{x})$.  To find this we
use Bayes' theorem (\ref{eq:bayesthm}), which we can write as a
proportionality

\begin{equation}
  \label{eq:bayesthm4}
p(\mu, \theta | \bvec{x}) \propto L(\bvec{x} | \mu, \theta) \pi(\mu, \theta) \;,
\end{equation}

\noindent where the prior pdf $\pi(\mu, \theta)$ specifies our degree
of belief in the parameters' values before carrying out the
measurement.

The problematic ingredient in the procedure above is the prior pdf
$\pi(\mu, \theta)$.  For a nuisance parameter $\theta$, one typically
has some specific information that constrains one's degree of belief
about its value.  For example, a calibration constant or background
event rate may be constrained by some control measurements, leading to
a best estimate $\tilde{\theta}$ and some measure of its uncertainty
$\sigma_{\theta}$.  Depending on the problem at hand one may from
these subsidiary measurements as well as physical or theoretical
constraints construct a prior pdf for $\theta$.  In many cases this
will be independent of the value of the parameter of interest $\mu$,
in which case the prior will factorize, i.e., $\pi(\mu, \theta) =
\pi_{\mu}(\mu) \pi_{\theta}(\theta)$.  For the present discussion we
will assume that this is the case.

The more controversial part of the procedure is the prior
$\pi_{\mu}(\mu)$ for the parameter of interest.  As one is carrying
out the measurement in order to learn about $\mu$, one usually does
not have much information about it beforehand, at least not much
relative to the amount one hopes to gain.  Therefore one may like to
write down a prior that is {\it non-informative}, i.e., it reflects a
maximal degree of prior ignorance about $\mu$, in the hopes that one
will in this way avoid injecting any bias into the result.  This turns
out to be impossible, or at least there is no unique way of
quantifying prior ignorance.

As a first attempt at a non-informative prior for $\mu$ we might choose
to take it very broad relative to the likelihood.  Suppose
as before that $\mu$ represents the rate of signal so we have
$\mu \ge 0$.  As an extreme example of a broad prior we may try

\begin{equation}
\label{eq:flatprior}
\pi_{\mu}(\mu)  = 
\left\{ \! \! \begin{array}{ll}
               1  & \quad \mu \ge 0  , \\*[0.2 cm]
               0 & \quad \mbox{otherwise}. 
              \end{array}
       \right.
\end{equation}

This so-called flat prior is problematic for a number of reasons.
First, it cannot be normalized to unit area, so it is not a proper
pdf; it is said to be {\it improper}.  Here this defect is not fatal
because in Bayes' theorem the prior always appears multiplied by the
likelihood, and if this falls off sufficiently rapidly as a function
of $\mu$, as is often the case in practice, then the posterior pdf for
$\mu$ may indeed be normalizable.

A further difficulty with a flat prior is that our inference is not
invariant under a change in parameter.  For example, if we were to
take as the parameter $\eta = \ln \mu$, then according to the rules
for transformation of variables we find for the pdf of $\eta$

\begin{equation}
\pi_{\eta}(\eta) = \pi_{\mu}(\mu) \left| \frac{ d\mu}{d \eta} \right|
= e^{\eta} \pi_{\mu}( \mu(\eta)) \;,
\end{equation}

\noindent so if $\pi_{\mu}(\mu)$ is constant then $\pi_{\eta}(\eta)
\propto e^{\eta}$ which is not.  So if we claim we know nothing about
$\mu$ and hence use for it a constant prior, we are implicitly saying
that we known something about $\eta$.

Finally we should note that the constant prior of
Eq.~(\ref{eq:flatprior}) cannot in any realistic sense reflect a
degree of belief, since it assigns a zero probability to the range
between any two finite limits.

The difficult and subjective nature of encoding personal knowledge
into priors has led to what is called {\it objective Bayesian
  statistics}, where prior probabilities are based not on an actual
degree of belief but rather derived from formal rules.  These give,
for example, priors which are invariant under a transformation of
parameters or which result in a maximum gain in information for a
given set of measurements.  For an extensive review see, for example,
Ref.~\cite{bib:KassWasserman96}; applications to HEP are discussed
in Refs.~\cite{bib:demortier,bib:casadei}.

The constant prior of Eq.~(\ref{eq:flatprior}) has been used in HEP so
widely that it serves a useful purpose as a benchmark, despite its
shortcomings.  Although interpretation of the posterior probability as
a degree of belief is no longer strictly true, one can simply regard
the resulting interval as a given function of the data, which will
with some probability contain the true value of the parameter.  Unlike
the confidence interval obtained from the frequentist procedure,
however, the coverage probability will depend in general on the true
(and unknown) value of the parameter.

We now turn to the Bayesian treatment of nuisance parameters.  What we
get from Bayes' theorem is the joint distribution of all of the
parameters in the problem, in this case both $\mu$ and $\theta$.
Because we are not interested in the nuisance parameter $\theta$ we
simply integrate (or sum in the case of a discrete parameter) to find
the marginal pdf for the parameter of interest, i.e.,

\begin{equation}
\label{eq:marginalize}
p(\mu | \bvec{x}) = \int p(\mu, \theta | \bvec{x}) \, d \theta \;.
\end{equation}

One typically has not one but many nuisance parameters and the
integral required to marginalize over them cannot be carried out in
closed form.  Even Monte Carlo integration based on the
acceptance-rejection method becomes impractical if the number of
parameters is too large, since then the acceptance rate becomes very
small.  In such cases, Markov Chain Monte Carlo (MCMC) provides an
effective means to calculate integrals of this type.  Here one
generates a correlated sequence of points in the full parameter space
and records the distribution of the parameter of interest, in effect
determining its marginal distribution.  An MCMC method widely
applicable to this sort of problem is the Metropolis-Hastings
algorithm, which is described briefly in Ref.~\cite{bib:Cowan09}.
In-depth treatments of MCMC can be found, for example, in the texts by
Robert and Casella~\cite{bib:Robert04}, Liu~\cite{bib:Liu01}, and the
review by Neal~\cite{bib:Neal93}.

\section{The Poisson counting experiment}
\label{sec:poislim}

As a simple example, consider an experiment in which one counts a
number of events $n$, modeled as following a Poisson distribution with
a mean of $s + b$, where $s$ and $b$ are the contributions from signal
and background processes, respectively.  Suppose that $b$ is known and
we want to test different hypothetical values of $s$.  Specifically,
we want to test the hypothesis of $s=0$ to see if we can establish the
existence of the signal, and regardless of whether we succeed in doing
this we can set an upper limit on $s$.

To establish discovery of the signal using the frequentist approach,
we test $s=0$ against the alternative of $s > 0$.  That is, we assume
the relevant signal models imply positive $s$, and therefore we take
the critical region of the test to correspond to larger numbers of
events.  Equivalently, we can define the $p$-value of the $s=0$
hypothesis to be the probability, assuming $s=0$, to find as many
events as actually observed or more, i.e.,

\begin{equation}
\label{eq:poissondisc}
p_0 = P (n \ge n_{\rm obs} | s=0, b) = \sum_{n = n_{\rm obs}}^{\infty}
\frac{b^n}{n!} e^{-b} \;.
\end{equation}

\noindent We can exploit a mathematical identity

\begin{equation}
\label{eq:poissonsum}
\sum_{n=0}^m P(n|b) = 1 - F_{\chi^2}(2b; n_{\rm dof})
\end{equation}

\noindent with $n_{\rm dof} = 2(m+1)$ to relate the sum of Poisson
probabilities in Eq.~(\ref{eq:poissondisc}) to the cumulative
chi-square distribution $F_{\chi^2}$, which allows us write the
  $p$-value as

\begin{equation}
\label{eq:poissonp0}
p_0 = F_{\chi^2}(2b; 2 n_{\rm obs}) \;.
\end{equation}

For example, suppose $b = 3.4$ and we observe $n_{\rm obs} = 16$
events.  Eq.~(\ref{eq:poissonp0}) gives $p_0 = 3.6 \times 10^{-6}$
corresponding to a significance $Z = 4.5$.  This would thus constitute
strong evidence in favour of a nonzero value of $s$, but is still
below the traditional threshold of $Z = 5$.

To construct the frequentist upper limit we should test all
hypothetical values of $s$ against to the alternative of $s=0$, so the
critical region consists of low values of $n$.  This means we take the
$p$-value of a hypothesized $s$ to be the probability to find $n$ as
small as observed or smaller, i.e.,

\begin{equation}
p_s = \sum_{m=0}^n \frac{(s+b)^m}{m!} e^{-(s+b)} \;.
\end{equation}

\noindent The upper limit at $\mbox{CL} = 1 - \alpha$ is found from
the value of $s$ such that the $p$-value is equal to $\alpha$, i.e.,

\begin{equation}
\alpha = \sum_{m=0}^n \frac{(s_{\rm up} +b)^m}{m!} e^{-(s_{\rm up} +b)}
= 1 - F_{\chi^2}\left[ 2(s_{\rm up} + b), 2(n+1) \right] \;,
\end{equation}

\noindent where in the second equality we again used the identity
(\ref{eq:poissonsum}) to relate the sum of Poisson probabilities to
the cumulative chi-square distribution.  This allows us to solve for
the upper limit

\begin{equation}
\label{eq:freqsuup}
s_{\rm up} = \frac{1}{2} F^{-1}_{\chi^2}\left[ 1 - \alpha, 2(n+1) \right] 
- b \;,
\end{equation}

\noindent where $F^{-1}_{\chi^2}$ is the chi-square quantile (inverse
of the cumulative distribution).  The upper limit $s_{\rm up}$ is
shown in Fig.~\ref{fig:poislim}(a) for $1 - \alpha = 95\%$ as a
function of $b$ for different numbers of observed events $n$.

\setlength{\unitlength}{1.0 cm}
\renewcommand{\baselinestretch}{0.9}
\begin{figure}[htbp]
\begin{picture}(10.0,5.)
\put(0.,0){\includegraphics[width=0.45\textwidth]{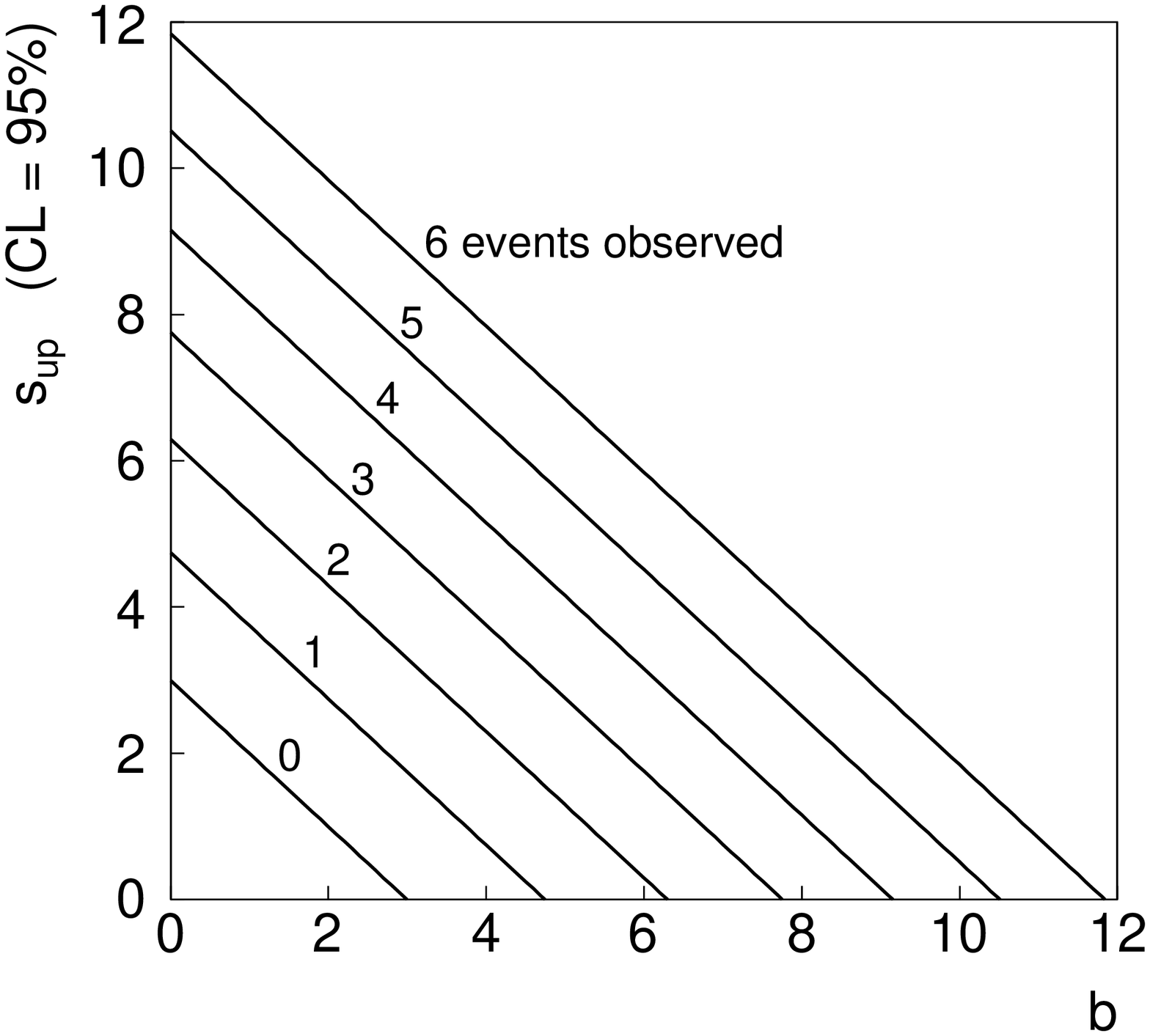}}
\put(6.,0){\includegraphics[width=0.45\textwidth]{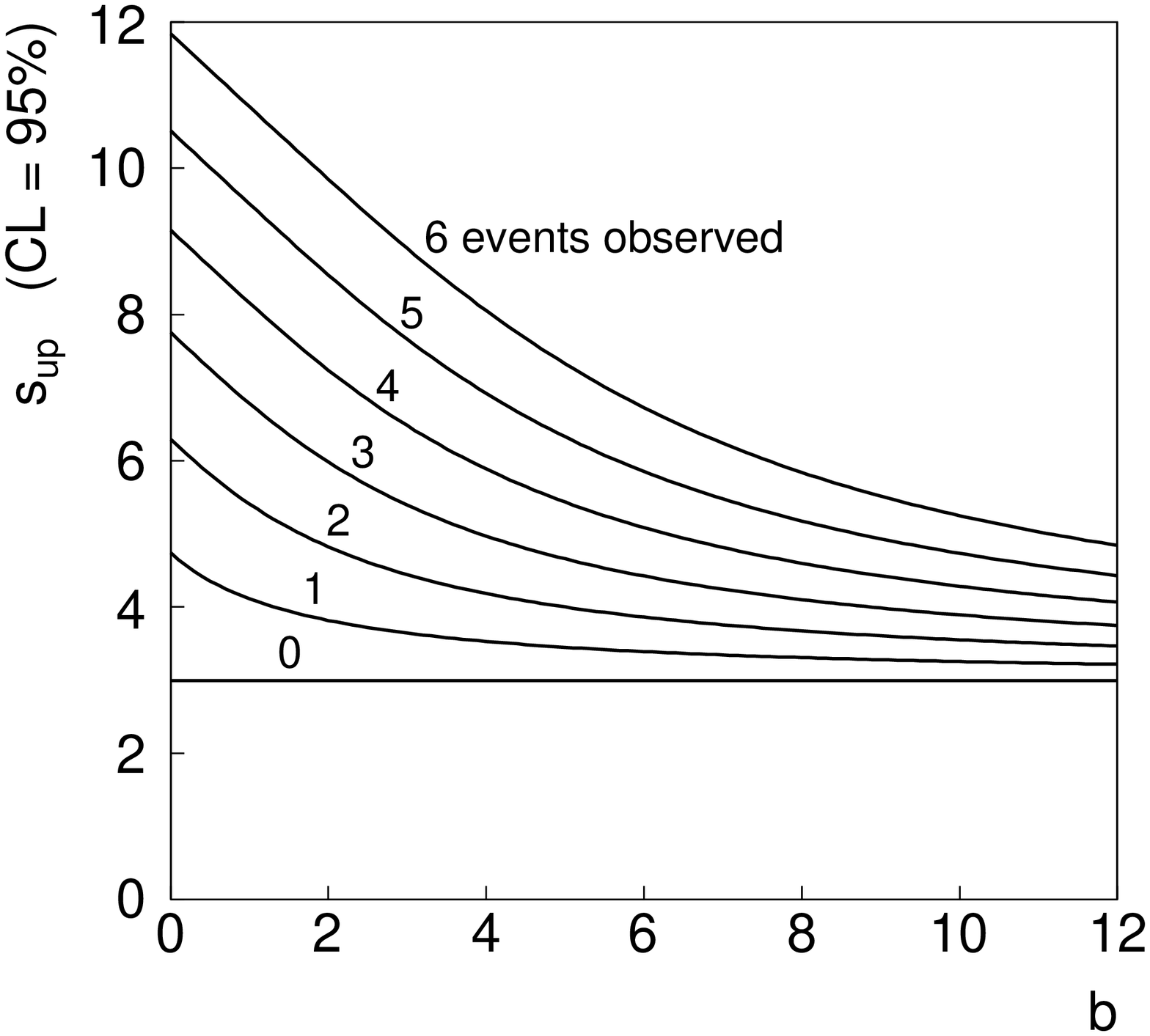}}
\put(5.2,4.5){(a)}
\put(11.3,4.5){(b)}
\end{picture}
\caption{\small Upper limits on the mean number of signal events $s$
  at 95\% confidence level as a function of the expected background
  $b$ for (a) the frequentist method and (b) Bayesian method with a
  flat prior.}
\label{fig:poislim}
\end{figure}
\renewcommand{\baselinestretch}{1}
\small\normalsize

To find the corresponding upper limit in the Bayesian approach we need
to assume a prior pdf for $s$.  If we use the flat prior of
Eq.~(\ref{eq:flatprior}), then by using Bayes' theorem we find the
posterior pdf

\begin{equation}
p(s | n ) \propto \frac{(s + b)^n}{n!} e^{-(s + b)} 
\end{equation}

\noindent for $s \ge 0$ and $p(\mu | n) = 0$ otherwise.  This can be
normalized to unit area, which gives 

\begin{equation}
p(s | n ) = \frac{(s + b)^n e^{-(s + b)}}{\Gamma(b, n+1)} \;,
\end{equation}

\noindent where $\Gamma(b, n+1) = \int_b^{\infty} x^n e^{-x} \, dx$
is the upper incomplete gamma function. 

Since in the Bayesian approach we are assigning a probability to
$s$, we can express an upper limit simply by integrating the
posterior pdf from the minimum value $s=0$ up to an upper limit
$s_{\rm up}$ such that this contains a fixed probability, say, $1 -
\alpha$.  That is, we require

\begin{equation}
1 - \alpha = \int_0^{s_{\rm up}} p(s | n) \, d s \;.
\end{equation}

\noindent To solve for $s_{\rm up}$ we can use
the integral

\begin{equation}
\int_0^a x^n e^{-x} \, dx = \Gamma(n+1) F_{\chi^2}(2a, 2(n+1)) \;,
\end{equation}

\noindent where again $F_{\chi^2}$ is the cumulative chi-square
distribution for $2(n+1)$ degrees of freedom.  Using this we find for
the upper limit

\begin{equation}
s_{\rm up} = \frac{1}{2} F^{-1}_{\chi^2} \left[ p, 2(n+1) \right]  - b \;,
\end{equation}

\noindent where

\begin{equation}
p = 1 - \alpha \left( 1 - F_{\chi^2}\left[ 2b, 2(n+1) \right] \right) \;.
\end{equation}

\noindent This is shown in Fig.~\ref{fig:poislim}(b).  Interestingly,
the upper limits for the case of $b=0$ happen to coincide exactly with
the values we found for the frequentist upper limit, and for nonzero
$b$ the Bayesian limits are everywhere higher.  This means that the
probability for the Bayesian interval to include the true value of $s$
is higher than $1 - \alpha$, so in this sense one can say that the
Bayesian limit is conservative.  The corresponding unified interval
from the procedure of Feldman-Cousins is described in
Ref.~\cite{bib:FC}.

If the parameter $b$ is not known, then this can be included in the
limit using the methods discussed above.  That is, one must treat $b$
as a nuisance parameter, and in general one would have some control
measurement that constrains its value.  In the frequentist approach
$b$ is eliminated by profiling; in the Bayesian case one requires a
prior pdf for $b$ and simply marginalizes the joint pdf of $s$ and $b$
to find the posterior $p(s|n)$.  The problem of a Poisson counting
experiment with additional nuisance parameters is discussed in detail
in Refs.~\cite{bib:demortier,bib:Cranmer03}.

\section{Limits in cases of low sensitivity}
\label{sec:spurious}

An important issue arises when setting frequentist limits that is
already apparent in the example from Sec.~\ref{sec:poislim}.  In
Fig.~\ref{fig:poislim}(a), which shows the frequentist upper limit on
the parameter $s$ as a function of $b$, one sees that $s_{\rm up}$ can
be arbitrarily small.  Naive application of Eq.~(\ref{eq:freqsuup})
can in fact result in a negative upper limit for what should be an
intrinsically positive quantity.  What this really means that all
values of $s$ are rejected in a test of size $\alpha$.  This can
happen if the number of observed events $n$ fluctuates substantially
below the expected background $b$.  One is then faced with the
prospect of not obtaining a useful upper limit as the outcome of one's
expensive experiment.  It might be hoped that such an occurrence would
be rare but by construction it should happen with probability
$\alpha$, e.g., 5\% of the time.

Essentially the same problem comes up whenever we test any hypothesis
to which we have very low sensitivity.  What ``low sensitivity'' means
here is that the distributions of whatever statistic we are using is
almost the same under assumption of the signal model being tested as
it is under the background-only hypothesis.  This type of situation is
illustrated in Fig.~\ref{fig:QdistNosens}(a), where here we have
labeled the model including signal $s+b$ (in our previous notation,
$\mu=1$) and the background-only model $b$ (i.e., $\mu=0$)).

\setlength{\unitlength}{1.0 cm}
\renewcommand{\baselinestretch}{0.9}
\begin{figure}[htbp]
\begin{picture}(10.0,5.)
\put(0.,0){\includegraphics[width=0.45\textwidth]{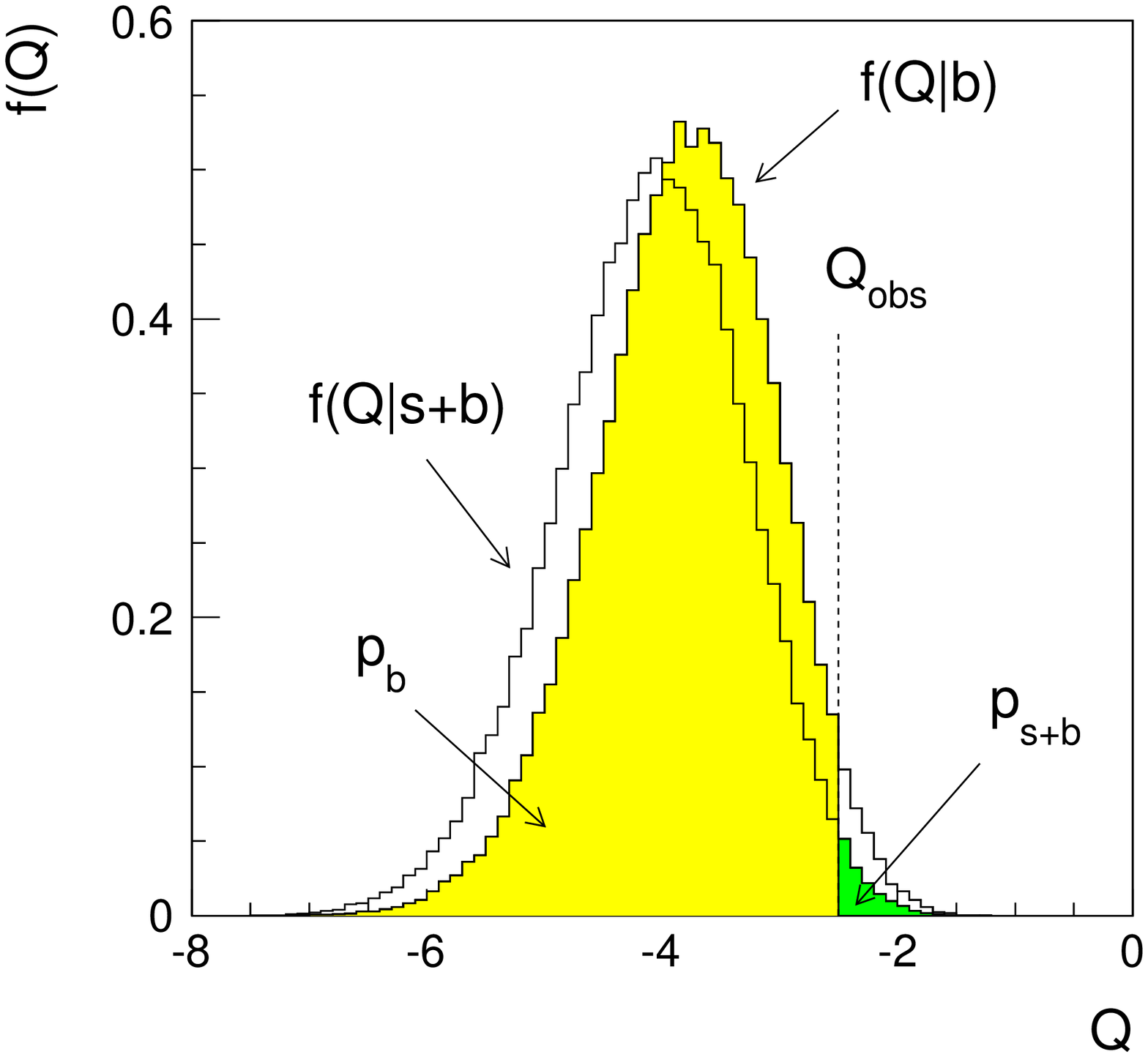}}
\put(6.,0){\includegraphics[width=0.45\textwidth]{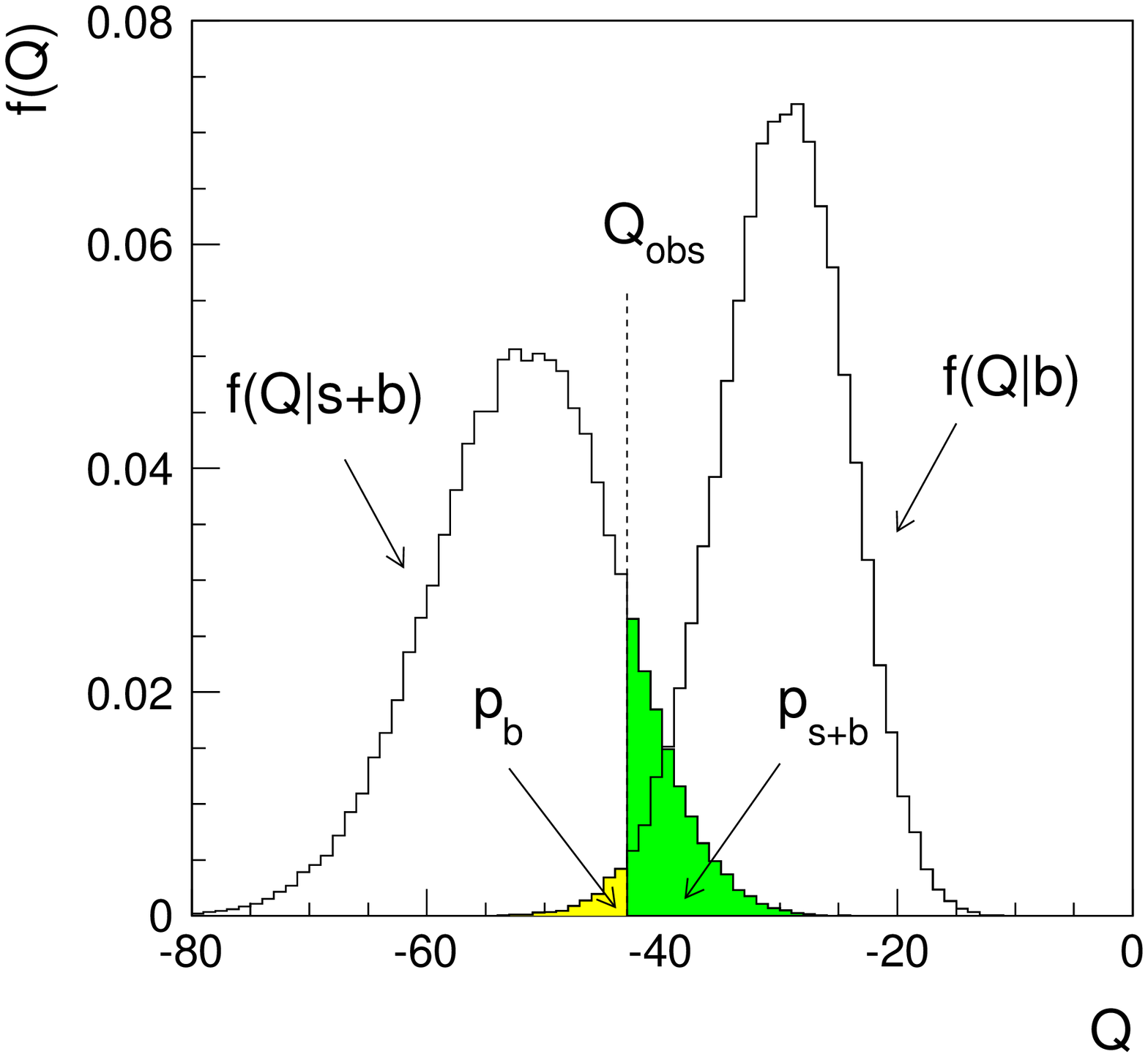}}
\put(5.2,4.5){(a)}
\put(11.2,4.5){(b)}
\end{picture}
\caption{\small (a) Distributions of the statistic $Q$ indicating low
  sensitivity to the hypothesized signal model; (b) illustration of
the ingredients for the CL$_{s}$ limit.}
\label{fig:QdistNosens}
\end{figure}
\renewcommand{\baselinestretch}{1}
\small\normalsize

The critical region for a test of the $s+b$ hypothesis consists of
high values of $Q$.  Equivalently, the $p$-value is the probability
$p_{s+b} = P(Q \ge Q_{\rm obs} | s + b)$.  Because the distributions
of $Q$ under both hypotheses are very close, the power of the test of
$s+b$ is only slightly greater than the size of the test $\alpha$,
which is equivalent to the statement that the quantity $1 - p_b$ is
only slightly greater than $p_{s+b}$.

If we have no sensitivity to a particular model, such as the
hypothesis of a Higgs boson with a mass much greater than what we
could produce in our experiment, then we do no want to not reject it,
since our measurement can produce no evidence to justify such a claim.
Unfortunately, the frequentist procedure that rejects the signal model
if its $p$-value is found less then $\alpha$ will do just that with a
probability of at least $\alpha$.  And this will happen even if the
model is, from an experimental standpoint, virtually indistinguishable
from the background-only hypothesis.  Since we typically take $\alpha
= 0.05$, we will exclude one model out of every twenty to which we
have no sensitivity.

One solution to this problem is the $\mbox{CL}_s$ procedure proposed
by Alex Read \cite{bib:read,bib:junk}, whereby the threshold for
rejecting a model is altered in a way that prevents one from rejecting
a model in the limit that one has very little sensitivity, but reverts
to the usual frequentist procedure when the sensitivity is high.  This
is achieved by defining

\begin{equation}
\label{eq:cls}
\mbox{CL}_s = \frac{P(Q \ge Q_{\rm obs} | s+b)}{P(Q \ge Q_{\rm obs} | b)} = 
\frac{p_{s+b}}{1 - p_b}  \;.
\end{equation}

\noindent The quantity $\mbox{CL}_s$ then is then used in place of the
$p$-value $p_{s+b}$, i.e., the $s+b$ model is rejected if one finds
$\mbox{CL}_s \le \alpha$.  The ingredients are illustrated in
Fig.~\ref{fig:QdistNosens}(b).

One can understand qualitatively how this achieves the desired goal by
considering the case where the distributions of $Q$ under the two
hypotheses $s+b$ and $b$ are close together.  Suppose the observed
value $Q_{\rm obs}$ is such that $p_{s+b}$ is less than $\alpha$, so
that in the usual frequentist procedure we would reject the $s+b$
hypothesis.  In the case of low sensitivity, however, the quantity $1
- p_b$ will also be small, as can be seen from
Fig.~\ref{fig:QdistNosens}(a).  Therefore the quantity $\mbox{CL}_s$
will be greater than $p_{s+b}$ such that the $s+b$ model is not
rejected by the criterion of Eq.~(\ref{eq:cls}).

If on the other hand the distributions are well separated, and $Q_{\rm
  obs}$ is such that the $p_{s+b} < \alpha$, then $p_b$ will also be
small and the term $1 - p_b$ that appears in the denominator of
$\mbox{CL}_s$ will be close to unity.  Therefore in the case with high
sensitivity, using $\mbox{CL}_s$ is similar to what is obtained from
the usual frequentist procedure based on the $p$-value $p_{s+b}$.

The largest value of $s$ not rejected by the $\mbox{CL}_s$ criterion
gives the corresponding $\mbox{CL}_s$ upper limit.  Here to follow the
traditional notation we have described it in terms of the mean number
of signal events $s$ rather than the strength parameter $\mu$, but it
is equivalent to using $\mbox{CL}_{\mu} = p_{\mu} / (1 - p_0)$ to
find an interval for $\mu$.

The $\mbox{CL}_s$ procedure described above assumes that the test
statistic $Q$ is continuous.  The recipe is slightly different if the
data are discrete, such as a Poisson distributed number of events $n$
with a mean $s+b$.  In this case the quantity $\mbox{CL}_s$ is defined
as

\begin{equation}
\label{eq:cls2}
\mbox{CL}_s = \frac{P(n \le n_{\rm obs} | s + b)}{P(n \le n_{\rm obs} | b)} \;,
\end{equation}

\noindent where $n_{\rm obs}$ is the number of events observed.  Here
the numerator is $p_{s+b}$, which the same as in Eq.~(\ref{eq:cls}).
The $p$-value of the background-only hypothesis is $p_b = P(n \ge
n_{\rm obs} | b)$, but the denominator in Eq.~(\ref{eq:cls2}) requires
$n$ less than {\it or equal} to $n_{\rm obs}$, so this is not exactly
the same as $1 - p_b$.  Equation~(\ref{eq:cls2}) is the fundamental
definition and it reduces to the ratio of $p$-values for the case of a
continuous test statistic.

For a Poisson distributed number of events, the $\mbox{CL}_s$ upper
limit coincides exactly with the Bayesian upper limit based on the
flat prior as shown in Fig.~\ref{fig:poislim}(b).  It is thus also
greater than or equal to the limit based on the $p$-value and is in
this sense conservative.  It also turns out that the $\mbox{CL}_s$ and
Bayesian limits (using a flat prior) agree for the important case of
Gaussian distributed data \cite{bib:read}.  The problem of exclusion
in the case of little or no sensitivity is mitigated in a different
way by the unified intervals seen in Sec.~\ref{sec:fc} by the
particular choice of the critical region (see, e.g.,
Ref.~\cite{bib:FC}).

\section{The look-elsewhere effect}
\label{sec:lee}

Recently there has been important progress made on the problem of
multiple testing, usually called in particle physics the
``look-elsewhere effect'' \cite{bib:lee,bib:lee2}.  The problem often
relates to finding a peak in a distribution when the peak's position
is not predicted in advance.  In the frequentist approach the correct
$p$-value of the no-peak hypothesis is the probability, assuming
background only, to find a peak as significant as the one found more
more so anywhere in the search region.  This can be substantially
higher than the probability to find a peak of equal or greater
significance in the particular place where it appeared.

The ``brute-force'' solution to this problem involves generating data
under the background-only hypothesis and for each data set, fitting a
peak of unknown position and recording a measure of its significance.
To establish a discovery one often requires a $p$-value less than $2.9
\times 10^{-7}$, corresponding to a $5\sigma$ effect.  Thus
determining this with Monte Carlo requires generating and fitting an
enormous number of experiments, perhaps several times $10^7$.  This is
particularly difficult in that under the background-only hypothesis
there is no real peak, but only fluctuations.  One of these
fluctuations will stand out as the most significant peak and this must
be found in order to determine the value of the test statistic such as
the profile likelihood ratio, $L(0)/L(\hat{\mu})$, for that particular
data set.  This must be repeated tens of millions of times without
failure of the fitting program, which is is a difficult computational
challenge.

In contrast, if the position of the peak were known in advance, then
the fit to the distribution would be much faster and easier, and
furthermore one can in many cases use formulae valid for sufficiently
large samples that bypass completely the need for Monte Carlo (see,
e.g., \cite{bib:asimov}).  But this ``fixed-position'' $p$-value would
not be correct in general, as it assumes the position of the peak was
known in advance.

Gross and Vitells \cite{bib:lee} have described a method that allows
one to modify the $p$-value computed under assumption of a fixed
position to obtain the correct value using a relatively simple
calculation.  Suppose a test statistic $q_0$, defined so that larger
values indicate increasing disagreement with the data, is observed to
have a value $u$.  Furthermore suppose the model contains a nuisance
parameter $\theta$ (such as the peak position) which is only defined
under the signal model (there is no peak in the background-only
model).  An approximation for the desired ``global'' $p$-value is
found to be

\begin{equation}
\label{eq:correctpval}
p_{\rm global} \approx p_{\rm local} + \langle N_u \rangle \;,
\end{equation}

\noindent where $p_{\rm local}$ is the $p$-value assuming a fixed
value of $\theta$ (e.g., fixed peak position), and $\langle N_u
\rangle$ is the mean number of ``upcrossings'' of the the statistic
$q_0$ above the level $u$ in the range of the nuisance parameter
considered (e.g., the mass range).

The value of $\langle N_u \rangle$ can be estimated from the number of
upcrossings $\langle N_{u_{0}} \rangle$ above some much lower value,
$u_0$, by using a relation due to Davis \cite{bib:davis87},

\begin{equation}
\label{eq:n1}
\langle N_u \rangle  \approx \langle N_{u_{0}} \rangle  e^{-(u-u_0)/2} \;.
\end{equation}

\noindent By choosing $u_0$ sufficiently low, the value of $\langle
N_u \rangle$ can be estimated by simulating only a very small number
of experiments, rather than the $10^7$ needed if one is dealing with a
$5\sigma$ effect.

Vitells and Gross also indicate how to extend the correction to the
case of more than one parameter, e.g., where one searches for a peak
of both unknown position and width, or for searching for a peak in a
two-dimensional space, such as an astrophysical measurement on the sky
\cite{bib:lee2}.  Here one may find some number of regions where
signal appears to be present, but within those regions there may be
islands or holes where the significance is lower.  In the
generalization to multiple dimensions, the number of upcrossings of
the test statistic $q_0$ is replaced by the expectation of a quantity
called the Euler characteristic, which is roughly speaking the number
of disconnected regions with significant signal minus the number of
`holes'.

It should be emphasized that an exact accounting of the look-elsewhere
effect requires that one specify where else one looked, e.g., the mass
range in which a peak was sought.  But this may be have been defined
in a somewhat arbitrary manner, and one might have included not only
the mass range but other variables that were also inspected for peaks
but where none was found.  It is therefore not worth expending great
effort on an exact treatment of the look-elsewhere effect, as would be
needed in the brute-force method mentioned above.  Rather, the more
easily obtained local $p$-value can be reported along with an
approximate correction to account for the range of measurements in
which the effect could have appeared.

\section{Examples from the Higgs search at the LHC}
\label{sec:higgs}

In this section we show how the methods described above have been
applied to the recent discovery of a Higgs-like boson at the LHC.  The
examples are taken from the analyses of the ATLAS experiment
\cite{bib:atlasHiggs}; similar results were obtained by CMS
\cite{bib:CMSHiggs}.

The Higgs search is more complicated than examples described earlier
because the production of something like a Higgs boson is
characterized by two parameters of interest: the strength parameter
$\mu$, which is defined here as the signal cross section divided by
the one predicted by the Standard Model, and the mass of the
resonance, here labeled $m_{\rm H}$.  The procedure has been to carry
out tests of $\mu$ for a set of fixed masses within a given range, and
the results are then interpolated.  One obtains from this two
important outputs, both as a function of $m_{\rm H}$: $p$-values for
the test of $\mu=0$ and confidence intervals (here, upper limits) for
$\mu$.

The $p$-value of the background-only hypothesis $p_0$ is shown versus
$m_{\rm H}$ in Fig.~\ref{fig:atlasp0}.  The values shown are not
corrected for the look-elsewhere effect; this is therefore referred to
as the {\it local} $p_0$.  On the right-hand side of the plot one can
see the value translated into the significance $Z$ according to
Eq.~(\ref{eq:significance}).  The lowest $p$-value is found at $m_{\rm
  H} = 126.5$ GeV and corresponds to $Z = 6.0$; taking into account
some additional systematic uncertainties in the electromagnetic energy
response reduces this to 5.9.

\setlength{\unitlength}{1.0 cm}
\renewcommand{\baselinestretch}{0.9}
\begin{figure}[htbp]
\begin{picture}(10.0,6)
\put(1.,0.){\includegraphics[width=0.7\textwidth]{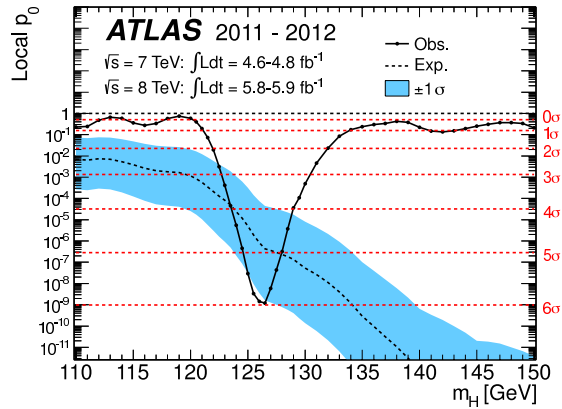}}
\end{picture}
\caption{\small The $p$-value of the background-only hypothesis
versus the Higgs mass $m_{\rm H}$ (from
  Ref.~\cite{bib:atlasHiggs}; see text).}
\label{fig:atlasp0}
\end{figure}
\renewcommand{\baselinestretch}{1}
\small\normalsize

The correction for the look-elsewhere effect is based on the procedure
described in Sec.~\ref{sec:lee} and in Ref.~\cite{bib:lee}.  If the
mass range of the search is taken to be 110 to 600 GeV, the peak
significance $Z$ reduces from 5.9 to 5.1; if one takes 110 to 150 GeV
it gives $Z= 5.3$.

The dotted line in Fig.~\ref{fig:atlasp0} gives the median value of
$Z$ under the hypothesis that the Higgs boson is present at the rate
predicted by the Standard Model, i.e., $\mu=1$.  That is, if one were
to generate a data set assuming an SM Higgs boson with a mass of 126.5
GeV, then this will lead to a certain significance $Z$ for a test of
$\mu=0$.  If one were to generate an ensemble of such experiments then
the median of the resulting distribution of $Z$ values, usually
referred to as the expected significance, is taken as a measure of the
sensitivity of the measurement.  The median $Z$ is preferred over the
expectation value because this is related to the median $p$-value
still through Eq.~(\ref{eq:significance}); for the expectation value
the corresponding relation would not hold.

For 126.5 GeV the expected significance is $Z = 4.9$, as can be seen
from the dotted line.  The blue band corresponds to the 68\%
inter-quantile range, i.e., the lower and upper edges of the band are
the 16\% and 84\% quantiles of the distribution (referred to as the
$\pm 1 \sigma$ band).  The band quantifies how much variation of the
result to expect as a result of statistical fluctuations if the
nominal signal model is correct.  From Fig.~\ref{fig:atlasp0} one can
see that the observed $p$-value is at the lower edge of the blue band.
So if the $\mu=1$ hypothesis is in fact correct, then the signal rate
observed by ATLAS fluctuated above the median value by a bit more than
one standard deviation.

Figure~\ref{fig:brazil} shows the upper limit on the strength
parameter $\mu$ as a function of the Higgs mass.  As with the case of
$\mu=0$ described above, the test procedure was carried out for a set
of discrete values of the mass and the results interpolated.  The
solid curve shows the observed upper limit using the CL$_s$ procedure
described in Sec.~\ref{sec:spurious}.  For each mass the distribution
of upper limits was found under assumption of background only, and the
dotted curve shows the median value.  The green and yellow bands show
the 68\% and 95\% inter-quantile ranges, i.e., the ranges that would
correspond to $\pm 1 \sigma$ and $\pm 2 \sigma$ if the distribution
were Gaussian.  In fact because the CL$_s$ procedure prevents one from
excluding very low values of $\mu$ the distribution of upper limits
can be significantly more asymmetric than a Gaussian.

\setlength{\unitlength}{1.0 cm}
\renewcommand{\baselinestretch}{0.9}
\begin{figure}[htbp]
\begin{picture}(10.0,5.5)
\put(1.,0.){\includegraphics[width=0.75\textwidth]{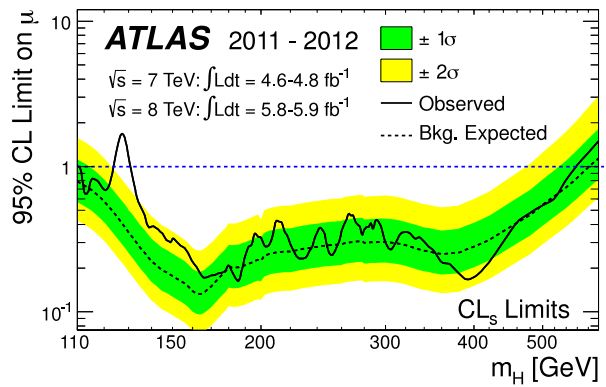}}
\end{picture}
\caption{\small CL$_s$ upper limits on the production cross section
  for the Higgs boson as a function of its mass (from
  Ref.~\cite{bib:atlasHiggs}; see text).}
\label{fig:brazil}
\end{figure}
\renewcommand{\baselinestretch}{1}
\small\normalsize

For almost all mass values the observed limit is close to the
expectation under assumption of $\mu=0$.  The exception is the mass
region around 126 GeV, where the upper limit is significantly higher.
This of course corresponds to the discovered signal.

\section{Why 5 $\sigma$?}
\label{sec:why5}

Common practice in HEP has been to regard an observed signal to be
worthy of the word ``discovery'' when its significance exceeds $Z =
5$, corresponding to a $p$-value of the background-only hypothesis of
$2.9 \times 10^{-7}$.  This is in stark contrast to many other fields
(e.g., medicine, psychology) in which a $p$-value of 5\% ($Z = 1.64$)
is considered significant.  In this section we examine critically some
of the reasons why the community has used such an extreme threshold.

First, it is not clear that the same significance threshold should be
used in all cases.  Whether one is convinced that a discovery is real
should take into account the plausibility of the implied signal and
how well it describes the data.  If the discovered phenomenon is a
priori very unlikely, then more evidence is required to produce a
given degree of belief that the new phenomenon exists.  As Carl Sagan
said, ``\ldots extraordinary claims require extraordinary evidence''
\cite{bib:sagan}.  This follows directly from Bayes' theorem
(\ref{eq:bayesthm4}), whereby the posterior probability of a
hypothesis is proportional to its prior probability.  If an
experimental result can only be explained by phenomena that may not be
impossible but nevertheless highly improbable (fifth force,
superluminal neutrinos), then it seems natural to demand a higher
level of statistical significance.

Some phenomena, on the other hand, are regarded by the community as
quite likely to exist before they are observed experimentally.  Most
particle physicists would have bet on the Higgs boson well in advance
of the direct experimental evidence.  As with the Higgs, however, when
a discovery is announced in HEP it is usually something fairly
important and the cost of an incorrect claim is perceived to be quite
high.  Every time the community endures a false discovery there is a
tendency to think that the threshold should be higher.

Another reason for the high five-sigma threshold is that the
experimenter may be unsure of the statistical model on which the
reported significance relies.  To first approximation one can think of
the significance $Z$ as the estimated size of the signal divided by
the standard deviation $\sigma$ in the estimated background.  Here
$\sigma$ characterizes the level of random fluctuation in the
background, i.e., it is a statistical error.  If we have a systematic
uncertainty in the background as well, then roughly speaking these
should get added in quadrature.  If an underestimate of our systematic
errors would result in our $\sigma$ being wrong by a factor of
several, then a mere three-sigma effect may be no real effect at all.
The high threshold in this case thus compensates for modeling
uncertainty.

Another important issue is the look-elsewhere effect, where as
discussed in Sec.~\ref{sec:lee} it is difficult to define exactly
where else one looked.  That is, should one correct for the fact that
the search histogram had 100 bins, or also for the fact that one
looked at 100 different histograms, or perhaps account for the
thousands of scientists all carrying out searches?  Surely in such a
scenario someone will see a bump in a histogram somewhere that appears
significant.  Since it is impossible to draw an unambiguous boundary
around where one ``looked'', there always remains a nagging feeling
that one's correction for this effect may have been inadequate, hence
the desire for a greater margin of safety before announcing a
discovery.

The $p$-value, however, really only addresses the issue of whether a
fluctuation in the background-only model is likely to lead to data as
dissimilar to background as what was actually obtained.  It is not
designed to compensate for systematic errors in the model, the cost of
announcing a false discovery or the plausibility of the phenomena
implied by the discovery.  Usually when a new phenomenon is
discovered, it appears initially as only marginally significant, then
continues to emerge until everyone is convinced.  At first, everyone
asks whether the apparent signal is just a fluctuation, but at some
point people stop asking that question, because it is obvious that
something has been observed.  The question is whether that something
is ``new physics'' or an uncontrolled systematic effect.  Provided
that the look-elsewhere effect is taken into account in a reasonable
way, this transition probably takes place closer to the three-sigma
level, in any case well before $Z=5$.

Nevertheless, the 5-sigma threshold continues to be used to decide
when the word ``discovery'' is appropriate.  In future the HEP
community should perhaps think of better ways of answering the
different questions that arise when searching for new phenomena, since
the statistical significance is really only designed to say whether
the data, in the absence of a signal, is likely to have fluctuated in
manner at least as extreme as what was observed.  Lumping all of the
issues mentioned above into the $p$-value simply makes them more
difficult to disentangle.

\section{Conclusions}
\label{sec:conclusions}

To discover a new physical phenomenon we need to be able to
demonstrate quantitatively that our data cannot be described using
only known processes.  In these lectures we have seen how statistical
tests allow us to carry out this task.  They provide a framework for
rejecting hypotheses on the basis that the data we observed were
uncharacteristic for them and more indicative of an alternative
explanation.  Frequentist statistical tests nevertheless prevent one
from asking directly certain seemingly relevant questions, such as
``what is the probability that my theory is true?''.  Bayesian
statistics does allow one to quantify such a degree of belief, at the
expense of having to supply subjective prior probabilities.  The
frequentist and Bayesian approaches answer different but related
questions and both are valuable tools.

We did not have time to discuss in detail many other statistical
issues such as Bayesian methods for establishing discovery,
multivariate techniques and more sophisticated means for improving the
accuracy of statistical models by introducing carefully motivated
nuisance parameters.  These methods will no doubt play an important
role when the LHC enters its next data-taking phase.

\begin{acknowledgement}
  I wish to convey my thanks to the students and organizers of the
  69th SUSSP in St.~Andrews for the stimulating environment, friendly
  atmosphere and lively discussions that resulted in a highly
  productive and enjoyable school.
\end{acknowledgement}

\input{biblio}

\end{document}

%% file: biblio.tex
%
%
%